\newcounter{myctr}

\documentclass{ws-acs}
\usepackage{url}
\usepackage{rotating}
\usepackage{graphicx,epsf,dsfont,amssymb}
\begin{document}

\makeatletter
\def\@biblabel#1{[#1]}
\makeatother

\markboth{B. Berche, C. von Ferber, T. Holovatch, Yu.
Holovatch}{Transportation Network Stability: a Case Study of City
Traffic}

%
\catchline{}{}{}{}{}
%

\title{TRANSPORTATION NETWORK STABILITY: A CASE STUDY OF CITY
TRANSIT}

\author{BERTRAND BERCHE}

\address{
Statistical Physics Group, P2M Dpt, Institut Jean Lamour, Nancy
Universit\'e, \\BP 70239, F-54506 Vand\oe uvre les Nancy, France\\
Bertrand.Berche@ijl.nancy-universite.fr}

\author{CHRISTIAN VON FERBER}

\address{Applied Mathematics Research Centre, Coventry
University,\\ Coventry CV1 5FB, UK. \&\\
Institut f\"ur Theoretische Physik II, Heintich-Heine Universit\"at
D\"usseldorf, \\
D-40225 D\"usseldorf, Germany\\
c.vonferber@coventry.ac.uk}

\author{TARAS HOLOVATCH}

\address{
Statistical Physics Group, P2M Dpt, Institut Jean Lamour, Nancy
Universit\'e, \\BP 70239, F-54506 Vand\oe uvre les Nancy, France, \&\\
Applied Mathematics Research Centre, Coventry
University,\\ Coventry CV1 5FB, UK\\
holovatch@gmail.com}

\author{YURIJ HOLOVATCH}

\address{
Institute for Condensed Matter Physics, National Academy of Sciences
of Ukraine,\\
1 Svientsitskii Str., 79011 Lviv, Ukraine\\
hol@icmp.lviv.ua}

\maketitle

\begin{history}
\received{(received date)}
\revised{(revised date)}
\end{history}

\begin{abstract}
The goals of this paper are  to present criteria, that  allow to
{\em a priori} quantify the attack stability of real world
correlated networks of finite size and to check how these
criteria correspond to  analytic results available for
infinite uncorrelated networks. As a case study, we consider public
transportation networks (PTN) of several major cities of the world.
To analyze their resilience against attacks either the network nodes
or edges are removed in specific sequences (attack scenarios).
During each scenario the size $S(c)$ of the largest remaining network
component is observed as function of the removed share $c$ of
nodes or edges.
To quantify the  PTN  stability with respect to different attack scenarios we
use the area below the curve described by $S(c)$ for $c\in[0,1]$
recently introduced (Schneider, C.~M, et al., \emph{PNAS}
\textbf{108} (2011) 3838) as a numerical measure of network robustness.
This measure captures the network reaction over the whole attack sequence.
 We present results of the analysis of PTN stability against node and
link-targeted attacks.
\end{abstract}


\section{Introduction} \label{I}

Taken the importance of transportation networks in different types of
natural and man-made structures the relevance
of their stability against disturbances be they individual failures or complete breakdown
is obvious. In turn, one may single out two main
ingredients which determine this stability, these are ({\em i})
dynamical features of transport processes that take part on such
networks, i.e. their fluctuating {\em load} and ({\em ii}) structural features
of the networks themselves, i.e. their {\em topology}
\cite{complex_networks}. Whereas a comprehensive treatment of
transportation network stability has to deal with both of these
mentioned factors, the complexity of the problem often calls
for a separate account and analysis of each factor.
Moreover, recently network structure stability has become the subject of a
separate field of research within complex network science, where the
{\em attack vulnerability} of a complex network is treated by means
of a combination of tools of random graph theory
\cite{complex_networks} and those of percolation theory
\cite{percolation} and statistical physics \cite{network_attacks}.
The very notion of attack vulnerability of a complex network
originates from earlier studies of computer networks and reflects the
decrease of network performance as caused by the removal or dysfunction of
either their nodes or links (or both) \cite{Albert00,Tu00}.

The study of network vulnerability against failure or attack has conceptually much
in common with studies of percolation and they gained a lot from
concepts and insights in percolation theory. However, standard percolation theory
\cite{percolation} deals with homogeneous lattices whereas the
non-homogeneity of complex networks gives rise to a variety of
phenomena which are particular for these structures. To give an
example, the empirical analysis of numerous scale-free real-world
networks (the www and the internet \cite{Albert00,Tu00}, metabolic
\cite{Jeong00}, food web \cite{Sole01}, protein \cite{Jeong01}
networks) has revealed that these networks display an unexpectedly high
degree of robustness under random failure. However, if the scenario
is changed towards ``targeted'' attacks, the same networks may appear to
be especially vulnerable \cite{Cohen00,Callaway00}. It is the
non-homogeneity of networks that allows to choose different attack
scenarios, i.e. to remove network links or nodes not at random, but
following specific sequences prepared according to characteristics
determining their `importance'. For vertex-targeted attacks, the sequence may be
ordered by decreasing vertex degree \cite{Barabasi99,Broder00} or
betweenness centrality \cite{Holme02} for the unperturbed
network and the attack successively removes vertices according to
this original sequence. One may further extend the above scenarios
by recalculating the characteristics of the remaining vertices after each removal step
and reordering the lists \cite{Albert00}. Former analysis has shown
that attacks according to recalculated lists often turn out to be
more effective \cite{Girvan02,Holme02}.

So far, the prevailing analytic results on complex network stability
have been obtained for idealized models of infinite networks. In
particular, important insight on network structure stability may
be gained assuming that a complex network may perform its function
as long as it possesses a giant connected component (GCC) i.e. a
connected subnetwork which in the limit of an infinite network
contains a finite fraction of the network. Under this assumption,
the network  robustness may be judged using the Molloy-Reed
criterion, which has been formulated for essentially treelike networks  with a given
node degree distribution $P(k)$ but otherwise random linking between
vertices. The criterion for a GCC to be present in such networks is
\cite{Cohen00,Callaway00,Molloy_Reed}:
\begin{equation}\label{1}
\langle k(k-2) \rangle \geq 0,
\end{equation}
where $\langle \dots \rangle$ means the ensemble average over
networks with given $P(k)$. Defining the Molloy-Reed parameter as
the ratio of the moments of the degree distribution
\begin{equation}\label{2}
\kappa^{(k)}= \langle k^2 \rangle/ \langle k \rangle,
\end{equation}
one may rewrite (\ref{1}) as:
\begin{equation}\label{3}
\kappa^{(k)}\geq 2.
\end{equation}
For an uncorrelated network the parameter $\kappa$ can be equally
represented by the ratio between the mean number $z_1$ of next neighbors
(which is by definition equal to the mean node degree
$\langle k \rangle$) and the mean number $z_2$ of second nearest neighbors:
\begin{equation}\label{4}
\kappa^{(z)}=z_2/z_1.
\end{equation}
In terms of $\kappa^{(z)}$, condition (\ref{3}) can be rewritten as:
\begin{equation}\label{5}
\kappa^{(z)}\geq 1.
\end{equation}

For obvious reasons, relations (\ref{3}), (\ref{5}) can not be
directly applied to real-world networks, which usually are
correlated and are of finite size. Therefore, an  important issue
which arises in the analysis of attack vulnerability of  real-world
networks is the choice  of the observables which may be used to measure
network stability. Since the GCC is well-defined only for an
infinite network, often the size of the largest network component
$S$ is used. Alternatively, one can estimate network
stability from the average shortest path lengths or their inverse values
\cite{Holme02,berche09}. Recently, a unique measure for robustness was introduced
\cite{schneider11a,schneider11b} and has been used to devise a
method to restructure a network and to make it more robust against a
malicious attack.
Observing the normalised size $S(c)$ of the largest component as function
of the share $c$ of removed vertices or links a measure of stability is provided
by the area A under the curve for the interval $c\in[0,1]$. We will normalise this
value as
\begin{equation}\label{5a}
A = 100\int_0^1 S(c) {\rm d}c.
\end{equation}
Here, the size of the largest component is normalised such that $S(0)=1$.
In this respect, the measure
captures the network reaction over the whole attack sequence. The goal of
this paper is to elaborate criteria, which  allow to give {\em a
priori} information on the attack stability of real world correlated networks
of finite size and to check how these criteria correspond to the
analytic results available for the infinite uncorrelated networks.
As a case study, we consider public transportation networks
(PTN) of several major cities of the world. This paper continues
studies initiated in \cite{berche09}, where we have considered PTN
attack vulnerability. The results presented below complement Ref.
\cite{berche09} by describing the effects of  link-targeted
attacks as well as by applying the above mentioned measure for network
robustness \cite{schneider11a,schneider11b} to evaluate attack
efficiency.

For the remaining part of the paper we will use the following set-up. In the
next section we will shortly describe our PTN database, attack
scenarios and the observables used to describe different features of
the PTNs considered here. Results for the transportation network stability against
node-targeted  and link-targeted  attacks will be given in sections
\ref{III} and \ref{IV}, correspondingly. In section \ref{V} we
present some observed correlations between PTN characteristics measured prior
to attack and the PTN stability  during attacks following different
scenarios. Discussions and outlook are presented in section
\ref{VI}.

\section{Database and attack scenarios description} \label{II}

The systematic analysis of PTN using tools of complex network theory
dates back to the early  2000-s \cite{PTN1} and continues to this day
\cite{berche09,Ferber09a,Ferber09b,Berche10,holovatch11,PTN2}.
 It has been revealed that these networks share common statistical
properties:  they appear to be strongly correlated small-world
structures with high values of clustering coefficients and
comparatively low mean shortest path values. The
power-law node degree distributions observed for many PTN give
strong evidence of correlations within these networks.

\begin{table}
\caption{Some characteristics of the PTNs analyzed in this study.
Types of transport taken into account: {\underline B}us, {\underline
E}lectric trolleybus, {\underline F}erry, {\underline S}ubway,
{\underline T}ram, {\underline U}rban train; $N$: number of
stations; $R$: number of routes.  The following characteristics are
given:  $\langle k \rangle$ (mean node degree); $\ell^{\rm max}$,
$\langle \ell \rangle$ (maximal and mean shortest path length); $C$
(relation of the mean clustering coefficient to that of the
classical random graph of equal size); $\kappa^{(z)}$,
$\kappa^{(k)}$ (c.f. Eqs. (\ref{4}), (\ref{2})); $\gamma$ (an
exponent in the power law (\ref{6}) fit, bracketed values indicate
less reliable fits, see the text). More data is given in
\cite{berche09,holovatch11}. \label{tab1}}
\begin{center}
\tabcolsep1.2mm
 {\small
\begin{tabular}{lrrrrrrrrrr}
\toprule City & Type &   $N$ & $R$ & $\langle k \rangle$ &
$\ell^{\rm max}$ & $\langle \ell \rangle$ & $C$ & $\kappa^{(z)}$ &
$\kappa^{(k)}$ &
$\gamma$ \\
\colrule
Berlin       &  BSTU  &  2992  &   211 &  2.58  &  68  &  18.5 &  52.8   & 1.96 & 3.16 & (4.30)   \\
Dallas       &  B     &  5366  &   117 &  2.18  & 156  &  52.0 &  55.0   & 1.28 & 2.35 & 5.49     \\
D\"usseldorf &  BST   &  1494  &   124 &  2.57  &  48  &  12.5 &  24.4   & 1.96 & 3.16 & 3.76     \\
Hamburg      &  BFSTU &  8084  &   708 &  2.65  & 156  &  39.7 &  254.7  & 1.85 & 3.26 & (4.74)   \\
Hong Kong    &  B     &  2024  &   321 &  3.59  &  60  &  11.0 &  60.3   & 3.24 & 5.34 & (2.99)   \\
Istanbul     &  BST   &  4043  &   414 &  2.30  & 131  &  29.7 &  41.0   & 1.54 & 2.69 & 4.04     \\
London       &  BST   &  10937 &   922 &  2.60  &  107 &  26.5 &  320.6  & 1.87 & 3.22 & 4.48     \\
Moscow       &  BEST  &  3569  &   679 &  3.32  &  27  &   7.0 &  127.4  & 6.25 & 7.91 & (3.22)   \\
Paris        &  BS    &  3728  &   251 &  3.73  &  28  &   6.4 &  78.5   & 5.32 & 6.93 & 2.62     \\
Rome         &  BT    &  3961  &   681 &  2.95  &  87  &  26.4 &  163.4  & 2.02 & 3.67 & (3.95)   \\
Sa\~o Paolo  &  B     &  7215  &   997 &  3.21  &  33  &  10.3 &  268.0  & 4.17 & 5.95 & 2.72     \\
Sydney       &  B     &  1978  &   596 &  3.33  &  34  &  12.3 &  82.9   & 2.54 & 4.37 & (4.03)   \\
Taipei       &  B     &  5311  &   389 &  3.12  &  74  &  20.9 &  186.2  & 2.42 & 4.02 & (3.74)   \\
\botrule
\end{tabular}
}
\end{center}
\end{table}

In this work we analyse a selection of PTNs drawing from a database
compiled by the present authors earlier and described in Refs.
\cite{berche09,Ferber09a,Ferber09b,Berche10,holovatch11}. The choice
for the selection of these PTNs is motivated by
the idea to collect network samples from cities of different
geographical, cultural, and economical background. Some characteristics
of these networks are given in table \ref{tab1}.
For each selected city the available information
on all different types of public transportation is included. More
data as well as details about the database are given in
\cite{berche09,holovatch11}. As one can see from the table, the typical number
of routes is several hundreds while the typical number of stops (i.e. network nodes)
is several thousands with a mean node degree of $\langle k \rangle \sim 3$. This number of
network nodes is to be related to comparatively low values for the mean and maximal shortest path.
As mentioned above, the node degree distribution $P(k)$ for some of the PTN
has been observed \cite{berche09,holovatch11} to display a power-law
decay
\begin{equation}\label{6}
P(k) \sim k^{-\gamma},
\end{equation}
for large values of the node degree $k$. Results for the corresponding exponent values
are given in the last column of table \ref{tab1}. If the distribution $P(k)$
is better fitted by an exponential decay, the exponent corresponding to a power-law
fit is given in brackets (this is the case for seven out of thirteen listed PTNs).
As a measure of local network correlation we give the mean clustering coefficient of each PTN
normalised by the value $C_{ER}$ for an Erdos-Renyi random
graph with the same numbers of nodes $N$ and links $M$, $C_{ER}=2M/N^2$. Recall that
the clustering coefficient $C(i)$ of a given node $(i)$ is the ratio of the number of links
$E_i$ between the $k_i$ nearest neighbours of node $(i)$ and the maximal possible number of
mutual links between these:
\begin{equation} \label{7}
C(i)  =\frac{2E_i}{k_i(k_i-1)}.
\end{equation}
The values of $C$ quoted in table \ref{tab1} give convincing evidence for the presence of strong
local correlations.

In our earlier work on the PTN resilience to attacks of different types we introduced
different scenarios to remove network nodes or links, to model random failure or attack. However,
the focus of that work was primarily on node-targeted attacks. In previous work
\cite{berche09,Ferber09b,Berche10}, the present authors have shown that for PTNs the most
effective attack scenarios correspond to removing nodes $(i)$ either with highest degree $k_i$ or
with highest betweenness centrality values ${\cal C}_B(i)$. For a given node $(i)$, the latter quantity
is defined as:
\begin{equation} \label{8}
 {\cal C}_B(i)= \sum_{j\neq i \neq k} \frac{\sigma_{jk}(i)}{\sigma_{jk}},
 \end{equation}
where $\sigma_{jk}$ is the number of shortest paths
between nodes $j$ and $k$  and $\sigma_{jk}(i)$ is the number of these paths
that go via node $(i)$.

The results presented below show the outcome of node- and link-targeted attacks,
where either nodes or links are removed following specific sequences corresponding
to so-called scenarios. For node-targeted attacks
we concentrate on five different scenarios, by selecting the nodes: (i)  at random, (ii)
according to their initial degree (prior to the attack) (iii) according to their degree recalculated after nodes of higher degree have been removed  (iv) according to their initial betweenness centrality (v) according to their recalculated betweenness centrality. The same five scenarios are implemented for
the  link-targeted attacks. However, in this case one has to generalize the notions
of node degree and betweenness centrality for links.
We will define the degree $k^{(l)}$ of the link between nodes $i$ and $j$ with degrees $k_i$ and $k_j$
as:
\begin{equation}\label{9}
k^{(l)}_{ij} = k_i + k_j - 2.
\end{equation}
For the simple graph with two vertices and a single link, the link degree will be zero, $k^{(l)}=0$,
while for any link in a connected graph with more than two
vertices the link degree will be at least one, $k^{(l)} \geqslant 1$.
The link betweenness centrality ${\cal C}_B^{(l)}(i)$ measures the
importance of a link $i$ with respect to the connectivity between
the nodes of the network. The link betweenness centrality is defined
as
\begin{equation}\label{10}
{\cal C}_{B}^{(l)}(i)= \sum_{s\neq t\in \cal{N}}
\frac{\sigma_{st}(i)}{\sigma_{st}},
\end{equation}
where $\sigma_{st}$ is the number of shortest paths between the two
nodes $s,t\in \cal{N}$, that belong to the network $\cal{N}$, and
$\sigma_{st}(i)$ is the number of shortest paths between nodes $s$
and $t$ that go through the link $i$.

The two subsequent sections demonstrate how PTNs react on attacks
of the above described scenarios when these attacks are targeted on PTN
nodes (section \ref{III}) and links (section \ref{IV}). To quantify the
outcome of these attacks we monitor the evolution of the normalised size $S(c)$ of the largest
network component as function of the share $c$ (with $0\leq c \leq 1$) of removed links or nodes:
\begin{equation}\label{11}
S(c)=N(c)/N(0),
\end{equation}
where $N(0)$ is the initial number of nodes of the largest connected component while $N(c)$ is the corresponding remaining number of nodes in that component after a share $c$ of nodes or links has been removed.
Obviously, any network of non-zero size will have a largest connected component.

\section{Node-targeted attacks} \label{III}

The outcome of attacks targeting PTNs nodes has been reported by the present authors
in Refs. \cite{berche09,Ferber09b,Berche10}. In particular,
results of attacks
of sixteen different scenarios have been presented and the most effective ones were  singled out. Here, we
recall the results of the five scenarios laid out in the previous
section. In particular, this will allow to compare these with the corresponding link-targeted attack
scenarios (section \ref{IV}) and analysing these to elaborate
criteria for network stability  (section \ref{V}).

\begin{figure}[th]
\centerline{\includegraphics[width=5.5cm]{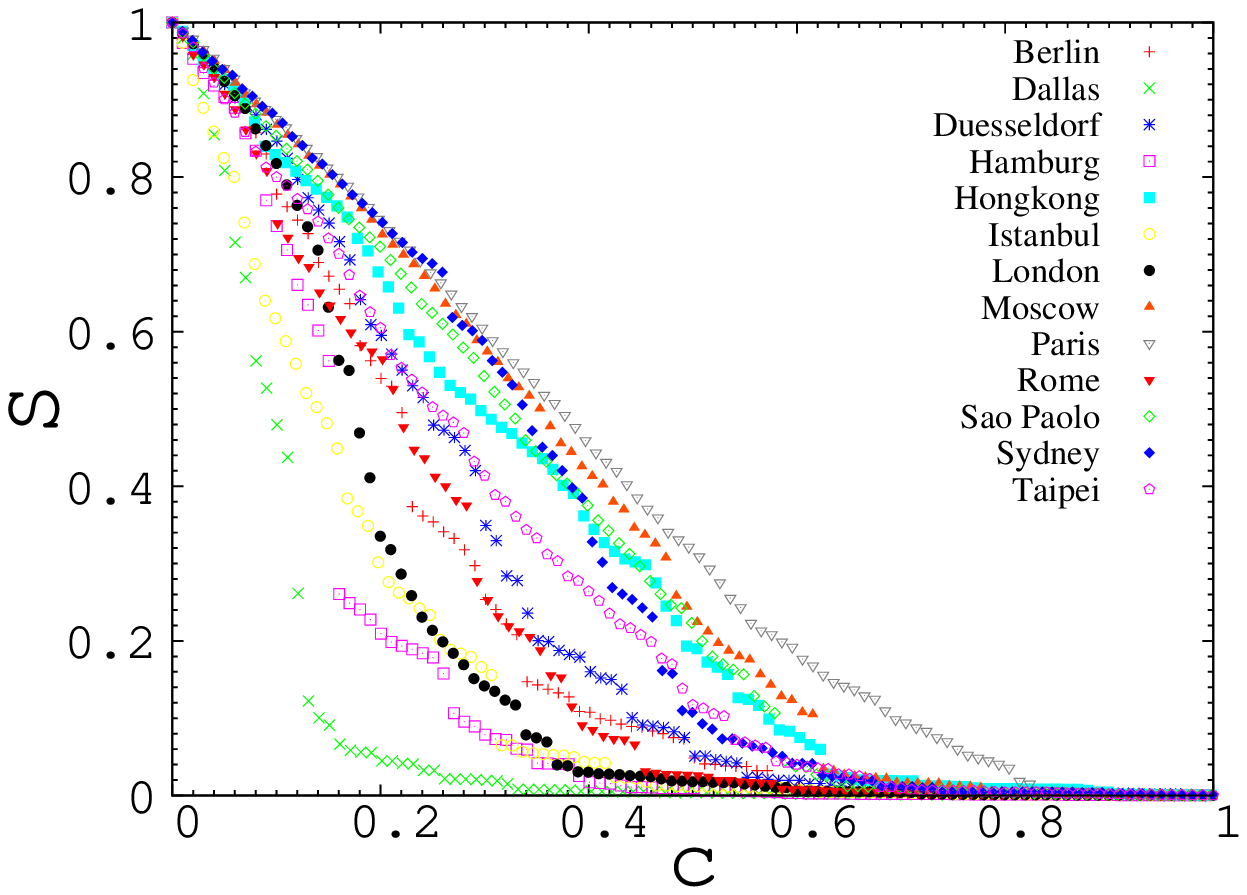} \hspace{3em}
\includegraphics[width=5.5cm]{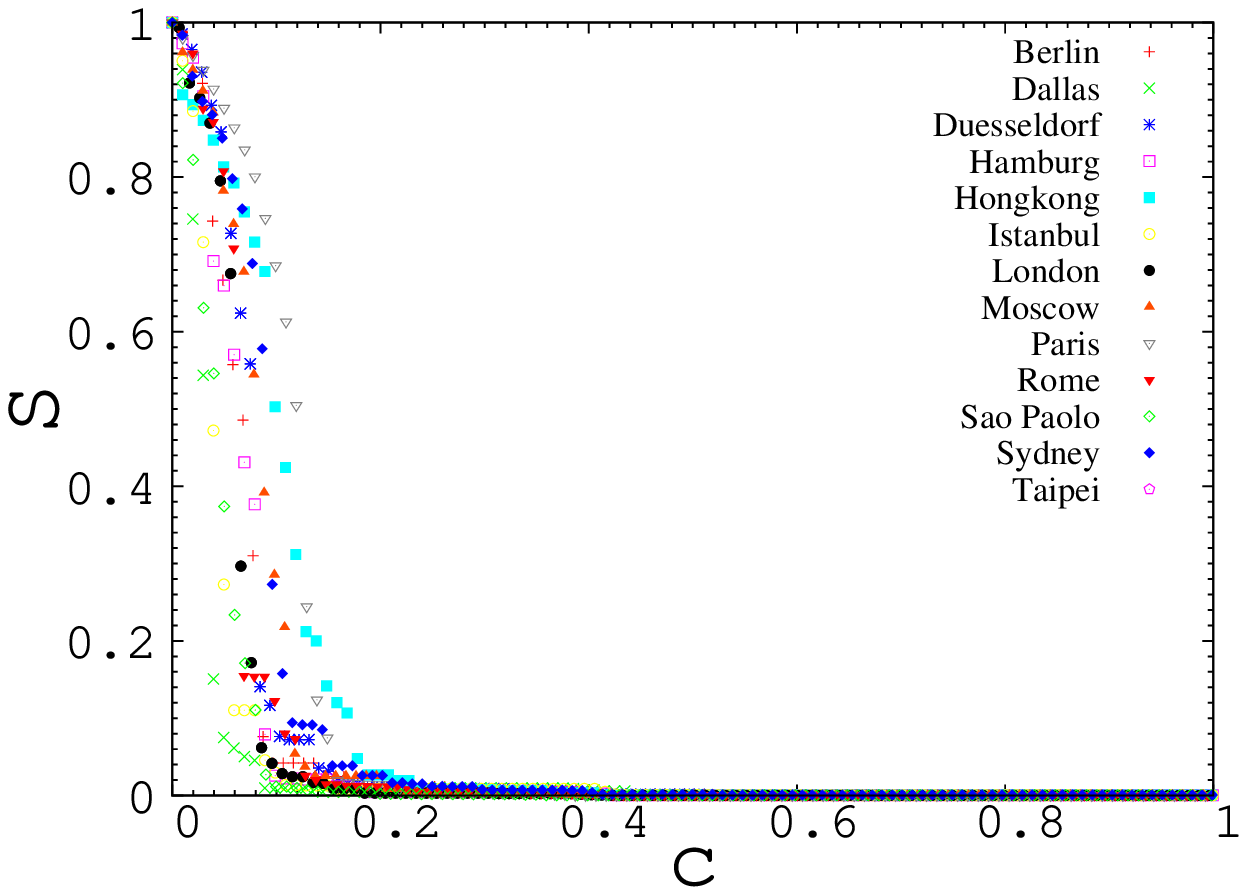}} \centerline{{\bf a}
\hspace{21em} {\bf b}}
 \vspace*{8pt}
\caption{Size of the largest cluster $S$ as functions of a fraction
of removed nodes $c$ normalized by their values at $c=0$. {\bf a}.
For random node-targeted scenario. {\bf b}. For recalculated
node-degree attack scenario. \label{fig1}}
\end{figure}

In figure \ref{fig1} we show the dependence of the normalized size
$S(c)$ (\ref{11}) of the largest connected cluster as function of
the share $c$ of removed PTN nodes for two attack scenarios: in the
first one, Fig. \ref{fig1} {\bf a} the PTN nodes are removed at
random, in the second one, Fig. \ref{fig1} {\bf b} the nodes are
removed according to the
 a list of the nodes ordered by their node degree $k$
recalculated after each step comprising the removal of 1\% of the
initial nodes. In the following, we will call this scenario the
'recalculated node degree scenario'. As noted, instead of
recalculating the PTN characteristics after the removal of each
individual node, the nodes are removed in groups of 1\% of the
initial nodes and the PTN characteristics are recalculated after the
removal of each such group. The random scenario Fig. \ref{fig1} {\bf
a} presents results of a single instance of an attack, we have
verified, however, that due to the large size of the PTNs size a
certain 'self-averaging' effect takes place: averaging of $S(c)$
over
 many random attack sequences instances do not significantly modify the
 picture for $S(c)$ presented in Fig. \ref{fig1} {\bf a}. As one
 may infer from the figures, the individual PTNs may react on the attacks in very
 different way, ranging from a gradual decrease of $S(c)$ as function of $c$
to sudden jumps at certain values of $c$.
A further striking feature of the plots visualising these scenarios is the qualitative
differences seen between individual PTNs as well as between different attack scenarios.

\begin{figure}[th]
\centerline{\includegraphics[width=5.5cm]{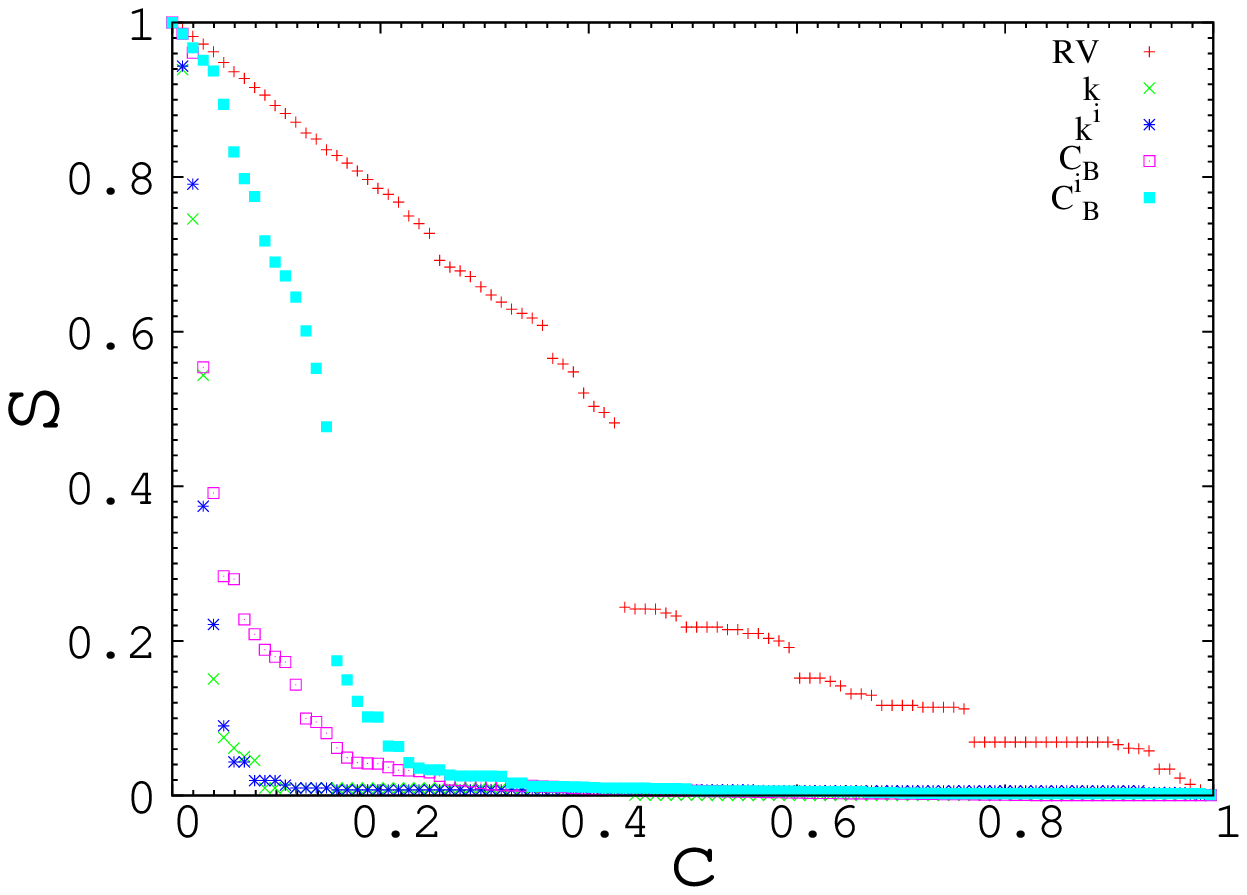} \hspace{3em}
\includegraphics[width=5.5cm]{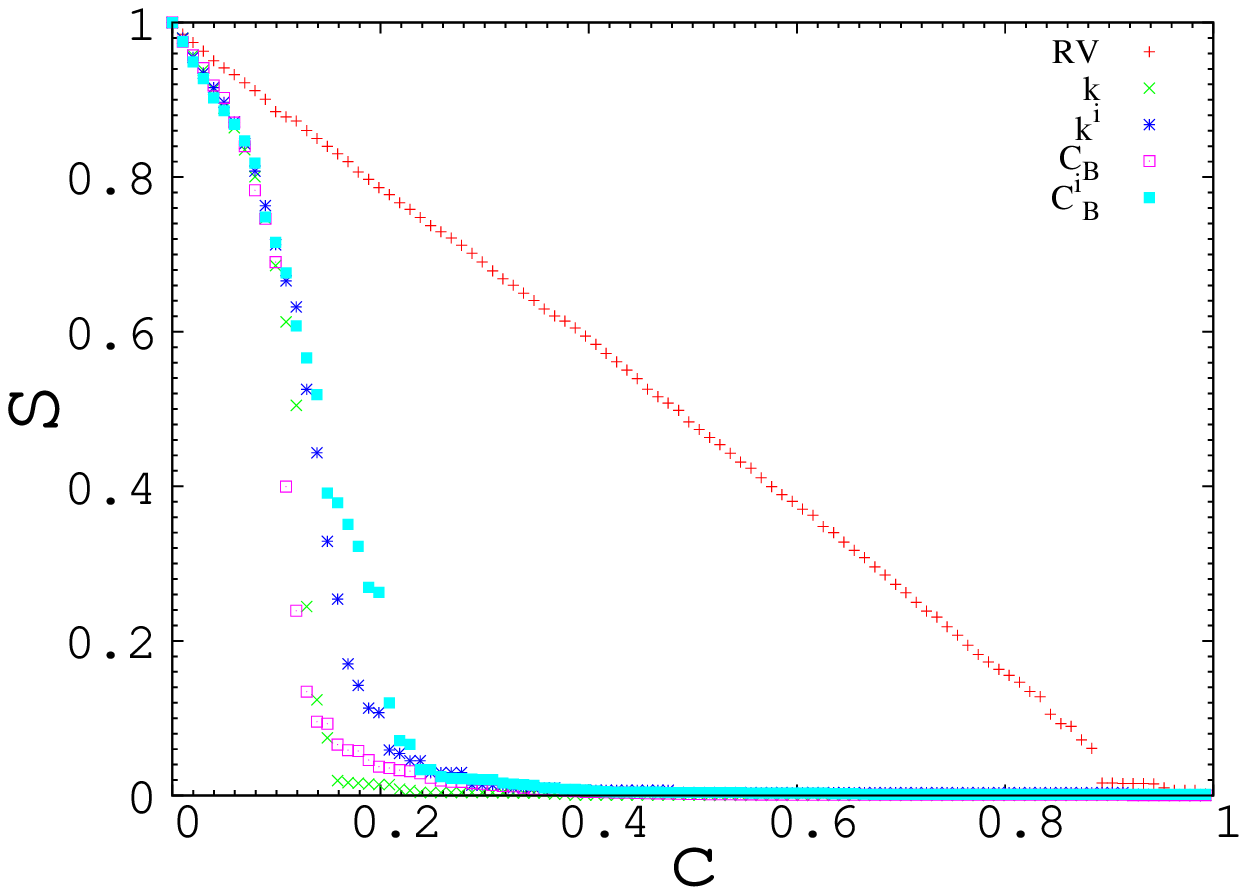}} \centerline{{\bf a}
\hspace{21em} {\bf b}}
 \vspace*{8pt}
\caption{The normalised largest component size $S(c)$ of the PTN as function of the
fraction $c$ of removed nodes for different attack scenarios. Each curve corresponds
to a different scenario defined by a corresponding sequence of nodes. RV: random vertex sequence;
$k$ and $k^i$: sequences ordered by recalculated and initial degrees; ${\cal C}_{B}$ and
 ${\cal C}^i_{B}$: sequences ordered by recalculated and initial betweenness.
{\bf a}. Five scenarios for the PTN of Dallas. {\bf b}. Five
scenarios for the PTN of Paris. \label{fig2}}
\end{figure}

To further illustrate the reaction of a given PTN to attacks of
different type, we present in Fig. \ref{fig2} the changes in the
largest component size of the PTNs of Dallas (Fig. \ref{fig2} {\bf
a}) and Paris (Fig. \ref{fig2} {\bf b}) for attacks of five
different scenarios, as described in the former section
\cite{Ferber09a}. For the case of the Paris PTN we observe that for
small values of the share $c$ of removed nodes ($c< 7\% $) these
scenarios cause practically indistinguishable impact on $S(c)$ and
$S(c)$ is a linear function of $c$. As $c$ increases, deviations
from the linear behavior arise and the impact of different scenarios
starts to vary. In particular, there appear differences between the
roles played by the nodes with highest value of $k$ and highest
betweenness centrality ${\cal C}_{B}$. Whereas the first quantity is
a local one, i.e. it is calculated from properties of the immediate
environment of each node, the second one is global. Moreover, the
$k$-based strategy aims to remove a maximal number of edges whereas
the ${\cal C}_{B}$-based strategy aims to cut as many shortest paths
as possible. In addition, there arise differences between the
'initial' and 'recalculated' scenarios, suggesting that the network
structure changes as important nodes are removed. Similar behavior
of $S(c)$ is observed for all PTNs included in this study, while
 the order of effectiveness of different attack scenarios may differ
between PTNs.

\section{Link-targeted attacks} \label{IV}

\begin{figure}[th]
\centerline{\includegraphics[width=5.5cm]{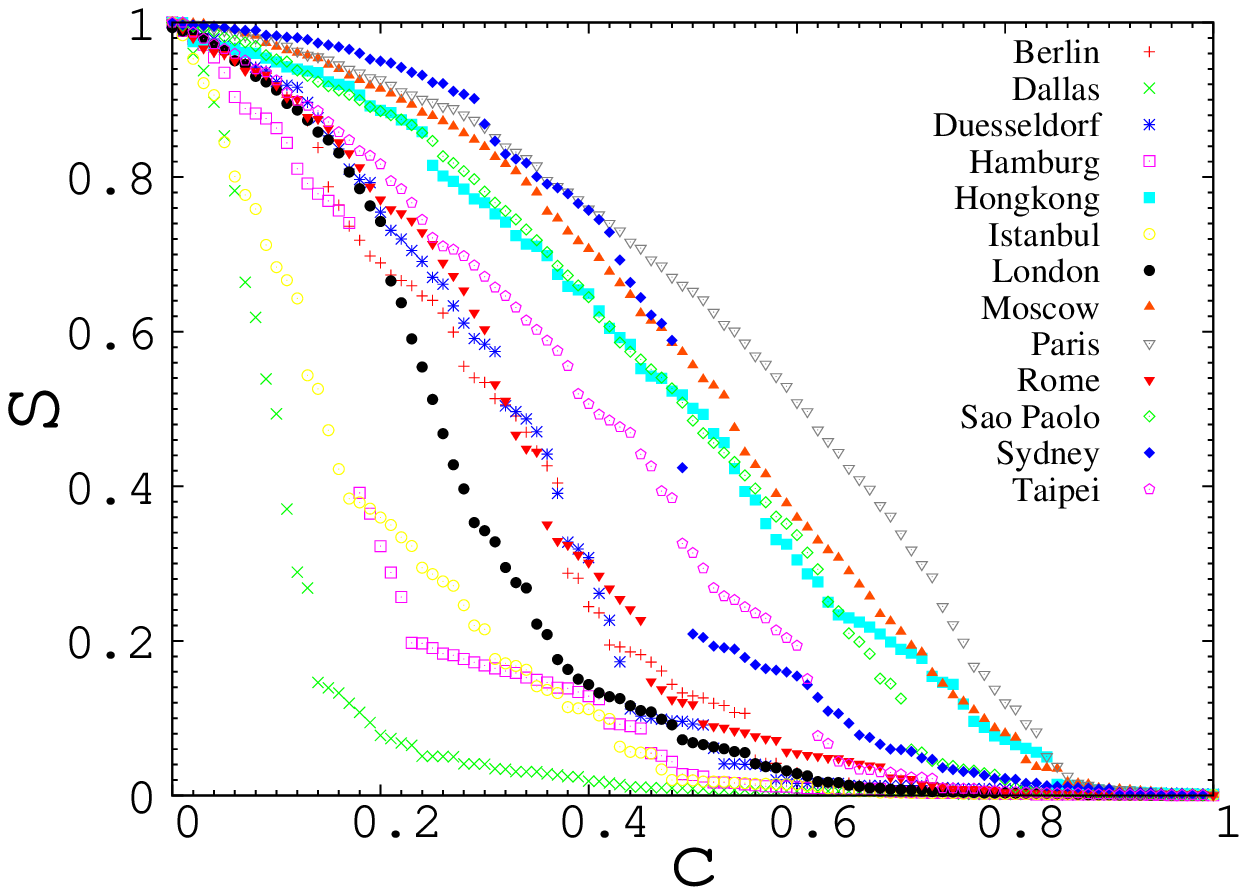} \hspace{3em}
\includegraphics[width=5.5cm]{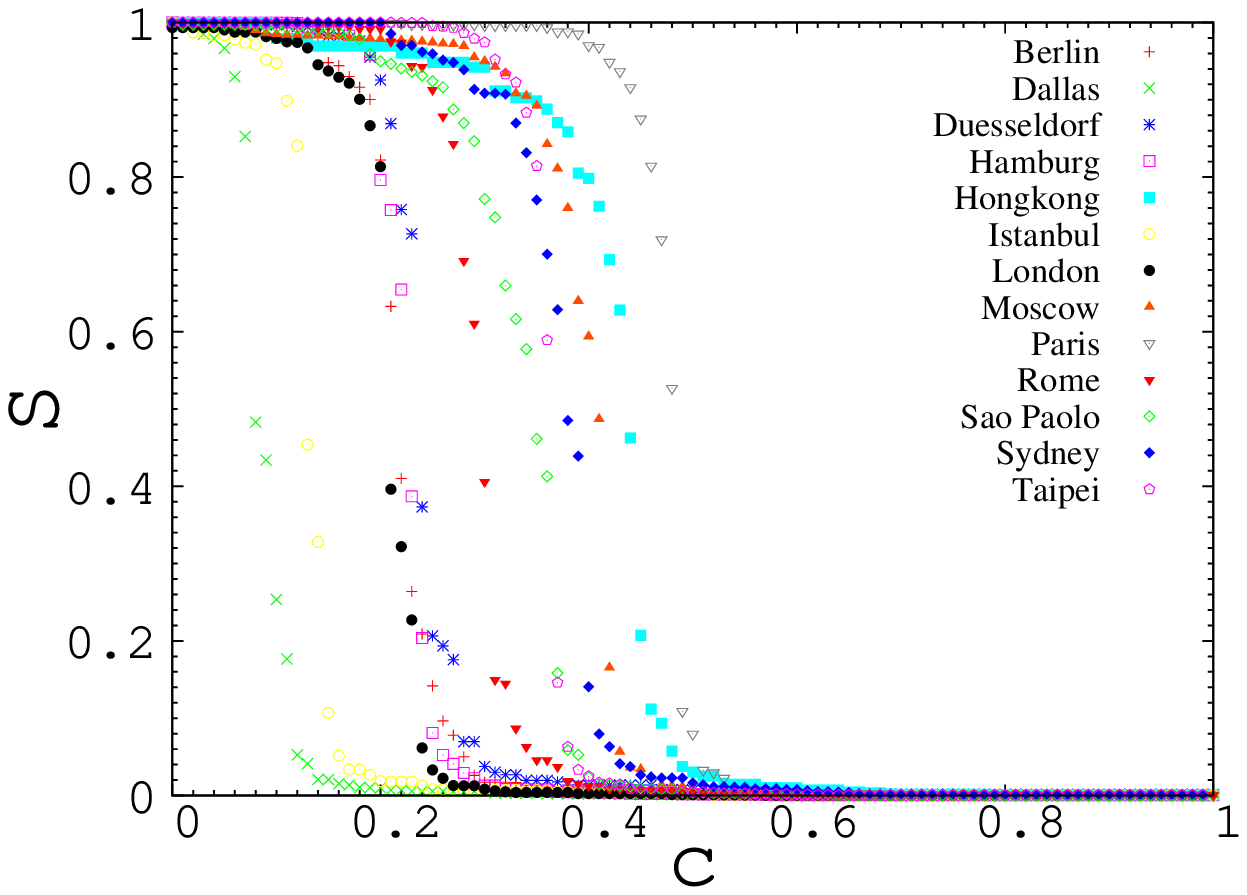}}\centerline{{\bf a}
\hspace{21em} {\bf b}}
 \vspace*{8pt}
\caption{The normalised size $S(c)$ of the largest cluster as
functions of the share of removed links for the PTNs of 13 cities.
{\bf a}. Random link-targeted scenario. {\bf b}. Recalculated
link-degree attack scenario. \label{fig3}}
\end{figure}

\begin{figure}[th]
\centerline{\includegraphics[width=5.5cm]{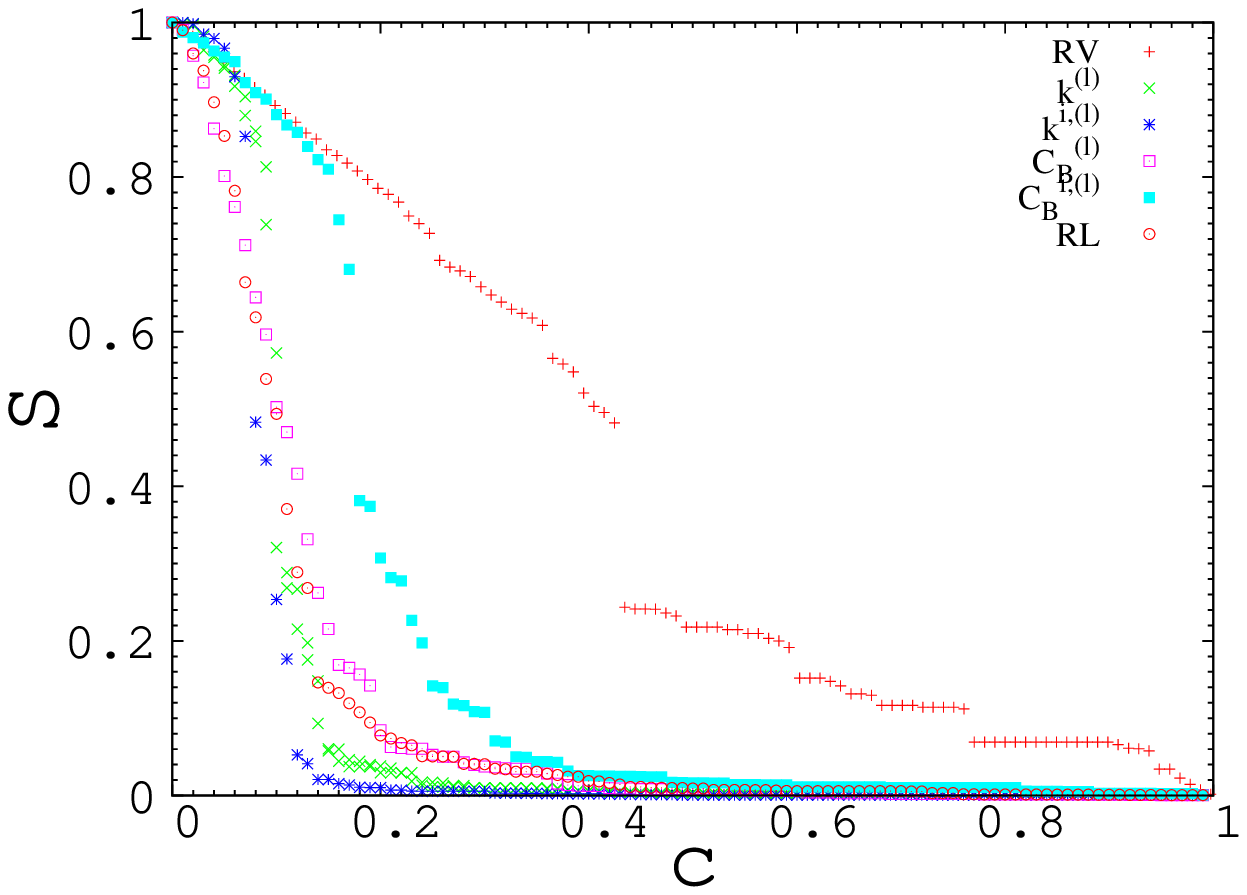} \hspace{3em}
\includegraphics[width=5.5cm]{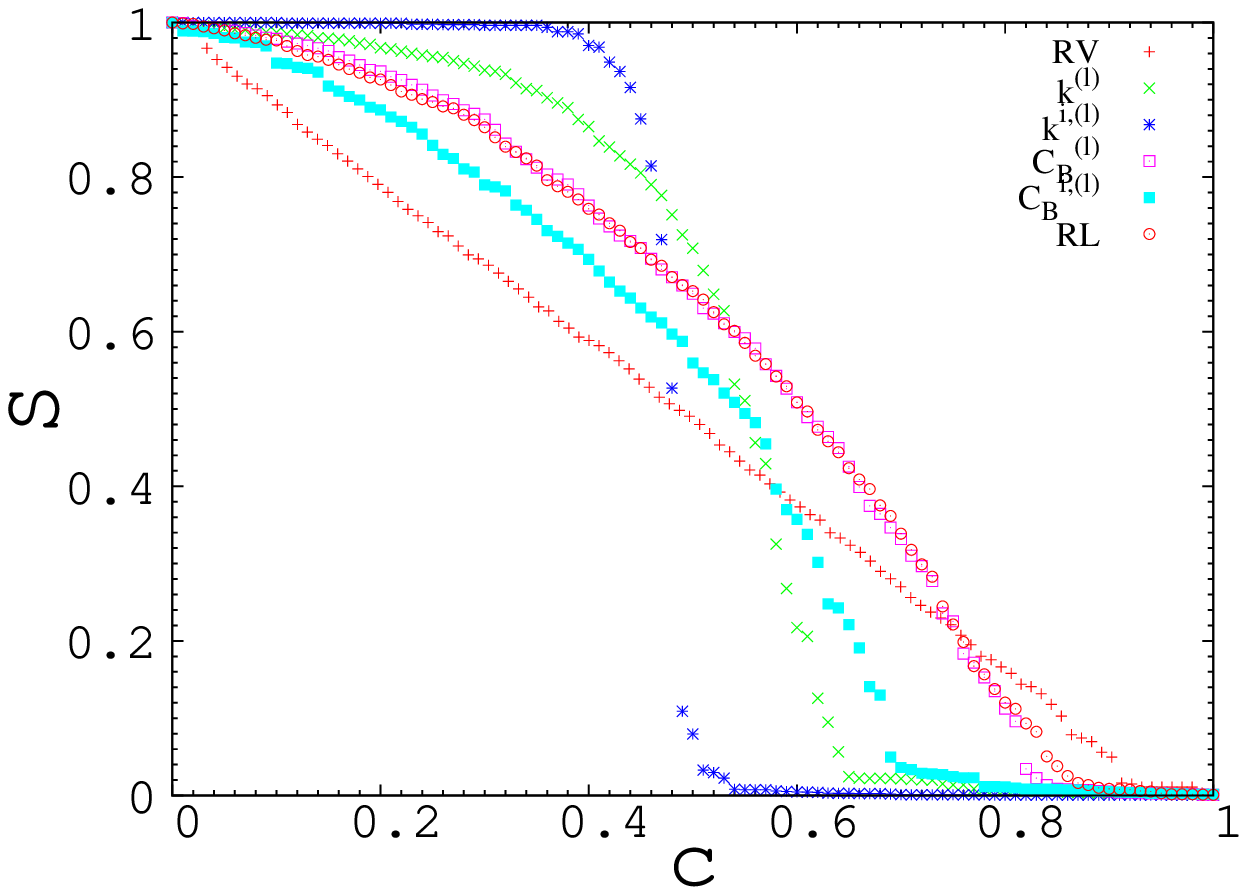}} \centerline{{\bf a}
\hspace{21em} {\bf b}}
 \vspace*{8pt}
\caption{Normalized size $S(c)$ of the largest component of the PTN as function of the share $c$ of
 removed links for different attack scenarios. Each curve corresponds to a different scenario as indicated
in the legend. Lists of removed links were prepared according to
their degree $k(l)$ and betweenness ${\cal C}_{B}(l)$ centrality. A
superscript $i$ refers to lists prepared for the initial PTN before
the attack; RL and RV denote the removal of a random link and
removal of random node respectively. {\bf a}. For PTN of Dallas.
{\bf b}. For PTN of Paris. \label{fig4}}
\end{figure}

A particular feature of link-targeted attacks is that when a link is
removed, the neighbouring nodes survive. Therefore, during the
link-targeted attacks all the nodes survive to the end of an attack,
i.e. the number of nodes does not change, while the share of the
removed links increases. In Figs. \ref{fig3} and \ref{fig4} we
monitor the behaviour of the normalised size  $S(c)$ for the largest
connected component (\ref{11}) but now as a function of the removed
links following corresponding link-attack scenarios. Besides
removing links at random we will use the sequences ordered according
to link degree and link betweenness centrality, (\ref{9}),
(\ref{10}) either calculated for the {\em initial} unperturbed PTN
(we will indicate the corresponding scenario by a superscript $i$,
e.g. ${\cal C}_{B}^{i,(l)}$) or following sequences with lists
recalculated for the remaining links after each step of removing 1\%
of the initial set of links.

In Fig. \ref{fig3} {\bf a} we show the change of the normalised size
$S(c)$ of the largest cluster under random link-targeted attacks
(RL). If one compares this behavior with that observed for the
random node removal scenario (RV) (see Fig. \ref{fig1} {\bf a}) one
can see, that for most PTNs with strong resilience to random
node-targeted attacks random link removal is even less effective. On
the other hand, for PTN with weak resilience there seems to be no
significant difference. Similar to the random node attacks (RV)
scenarios the random link attacks (RL) lead to changes of the
largest connected component $S$ that range from an abrupt breakdown
(Dallas) to a slow smooth decrease (Paris). The decay is even slower
than for random node removal - removing a link does not necessary
lead to removing a node from the largest cluster, while removing a
node from the completely connected network decreases it at least by
one node.

Typical results for PTNs under different types of link-targeted
attacks as applied to the PTNs of Dallas and Paris are displayed in
Fig. \ref{fig4}. We show how the normalised size $S(c)$ of the
largest connected component
 of the Dallas ({\bf a}) and Paris ({\bf b}) PTN varies as function of the
share $c$ of removed links following the
above described attack scenarios. As one can see,
 there is no significant difference between the effectiveness
of most scenarios including the random one  for the PTN of
Dallas. The vulnerability behavior
of the Dallas PTN under link-targeted attacks appears not to differ from
the corresponding random vertex removal approach. For Paris the situation is
quite different. The main observation is that initially the random vertex attack is
more effective than any link-targeted attack,
until breakdown and further, and only once the near to
50\% of the links have been removed the recalculated link
degree ($k(l)$) targeted scenario starts to be more harmful.
Comparing different link-targeted scenarios one notices
similar behavior between these, only the recalculated degree scenario line
initially decays slower, however to become more effective near to the
breakdown. In the following section we will compare outcomes of
node- and link-targeted  attacks in more detail.

The special behaviour of the recalculated link degree behaviour may be explained
as follows: in each removal step the links with highest link degrees are removed, However,
these may belong to a number of different nodes. The affected nodes will therefore remain,
however, with lower degrees. After recalculation the links on these nodes affected in the last step
will have moved to lower places in the ordered list such that other links will be affected. This will continue until
the degrees of all nodes have been reduced to three or less. At that point the removal of almost any link will
cut down the connected component and a rapid breakdown of the largest component takes place.

When the sequence for the removal of links is calculated using the initial degrees of the vertices,
then obviously in the first 1\% step a set of all links connected to the highest degree
vertices are removed which
is approximately equivalent to removing the corresponding 1\% of all highest degree nodes. As far as no
recalculation is involved the second step will essentially cut the links off the second 1\% of highest degree nodes.
Disregarding correlations between these operations one may therefore expect that the initial link and node degree
scenarios result in `similar breakdown behaviour.

\section{Robustness measures and correlations} \label{V}

\begin{table}
\caption{Robustness measure $A$, Eq. (\ref{5a}), for the PTNs of different cities
as analyzed in this study. Columns 2-6 give the value of $A$ for
 node-targeted attacks, columns 7-11 give $A$ for
link-targeted attacks. The results for $A$ for the following attack
scenarios are reported --  RV: random node; $k$ : node with
maximal recalculated degree; $k^i$ : node with maximal initial degree;
 ${\cal C}_{B}$ : node with maximal recalculated betweenness centrality;
${\cal C}_{B}^{i}$ : node with maximal initial betweenness centrality;
  RL : random link; $k^{(l)}$ : link with recalculated maximal degree;
$k^{i,(l)}$ : link with maximal initial degree;
${\cal C}_{B}^{(l)}$ : link with maximal recalculated betweenness;
${\cal C}_{B}^{i,(l)}$ : link with maximal initial betweenness \label{tab2}}
\begin{center}
\tabcolsep1.2mm
 {\small
\begin{tabular}{lrrrrrrrrrr}
\toprule
\multicolumn{1}{c}{City}&
\multicolumn{5}{c}{Node-targeted attacks}&
\multicolumn{5}{c}{Link-targeted attacks}\\
& RV & $k$ & $k^i$ & ${\cal C}_{B}$ & ${\cal C}_{B}^{i}$ &RL& $k^{(l)}$ & $k^{i,(l)}$ &
 ${\cal C}_{B}^{(l)}$ & ${\cal C}_{B}^{i,(l)}$  \\
\colrule
Berlin       &  22.71 &   6.52 &   7.12 &    7.27 &    9.44 &     31.21 &   22.27 &   25.57 &   29.91 &   30.92\\
Dallas       &   9.81 &   3.41 &   3.61 &    6.07 &   13.28 &     11.17 &   8.94  &   10.68 &   11.75 &   19.58\\
D\"uesseldorf&  25.47 &   7.45 &   9.39 &    8.26 &   12.65 &     31.22 &   23.88 &   28.69 &   30.58 &   31.44\\
Hamburg      &  15.82 &   6.34 &   6.99 &    6.53 &   12.19 &     20.74 &   22.49 &   24.02 &   20.22 &   20.47\\
Hong Kong    &  31.57 &   9.99 &   9.78 &    6.1  &   15.0  &     47.55 &   41.41 &   40.17 &   47.08 &   34.13\\
Istanbul     &  16.05 &   4.46 &   5.03 &    5.62 &    9.42 &     18.45 &   13.13 &   15.1  &   19.78 &   18.86\\
London       &  29.31 &   5.45 &   6.28 &    8.71 &   14.17 &     27.45 &   20.95 &   22.85 &   27.2  &   27.33\\
Moscow       &  34.61 &   8.02 &   8.37 &    7.82 &   11.63 &     51.18 &   38.99 &   41.96 &   50.68 &   41.58\\
Paris        &  37.93 &  10.77 &  13.12 &   10.67 &   14.07 &     56.04 &   47.12 &   51.83 &   55.93 &   48.03\\
Rome         &  22.26 &   6.61 &   7.68 &    7.05 &   14.81 &     32.52 &   29.2  &   27.8  &   33.99 &   30.13\\
S\~aopaolo   &  32.4  &   4.43 &   4.59 &    5.22 &    6.23 &     47.09 &   33.19 &   32.08 &   47.46 &   33.85\\
Sydney       &  32.15 &   8.74 &   9.49 &    6.61 &   18.53 &     46.45 &   37.26 &   35.74 &   49.14 &   26.15\\
Taipei       &  27.59 &  10.92 &  13.55 &   11.71 &   20.31 &     39.35 &   36.03 &   40.41 &   38.21 &   35.37\\

\botrule
\end{tabular}
}
\end{center}
\end{table}

As it was mentioned in the Introduction, different indicators may be
used in order to evaluate network stability. Here, for this purpose
we will use a measure, recently introduced in Refs.
\cite{schneider11a,schneider11b}. In our case, this measure
corresponds to the area below the curve describing the normlised
size $S(c)$ as function of the share $c$ of removed links, as
defined by Eq. (\ref{5a}).  As  follows from the definition, the
measure captures the effects on the network over the complete attack
sequence. It is especially useful in the analysis of the real-world
networks which are of finite size and usually are not characterized
by a single well-defined concentration at which phenomena analogous
to percolation (network clustering) occurs. Instead, the value $A$
is an integral characteristics, which is well-defined for a
finite-size network and is, as we will see below, nicely suited to
compare robustness of different PTN during attacks. In table
\ref{tab2} we give the value of $A$ for the node- and link-targeted
attacks (left and right parts of the table, correspondingly).
Columns marked as RV (RL) give $A$ for the attacks at which nodes
(links) were chosen at random, these numbers can be compared with
the outcome of attacks made according to the initially prepared
sequences of nodes (links) ordered by decreasing degrees ($k^i$,
$k^{i,(l)}$) and betweenness centralities (${\cal C}^i_B$, ${\cal
C}_B^{i,(l)}$). For the last four scenarios these indicators were
recalculated after each step of the attack, and the corresponding
results are given in columns marked as $k$, $k^{(l)}$ and ${\cal
C}_B$, ${\cal C}_B^{(l)}$.

With the data of table \ref{tab2} at hand, it is easy to compare the robustness of
a given PTN to attacks of different scenarios as well as to compare the robustness
of different PTNs. Assuming that the most stable PTNs are those characterized
by larger values of $A$ one may conclude from the table, that for the node-targeted attacks the
most harmful appear to be attacks targeted either on the nodes of highest degree (PTN of
Berlin, Dallas, D\"usseldorf, Hamburg, Istanbul, London, Rome, Sa\~opaolo, and Taipei)
or on the nodes of highest betweenness centrality (Hong Kong, Moscow, Paris, Sydney).
Another observation is that attacks performed according to the lists of
nodes recalculated after each step of the attack scenario appear to be more effective than
those performed according to the lists prepared prior to the attack. Moreover,
this difference is much more pronounced for the highest betweenness centrality
targeted nodes as for those with highest node degree.

On the other hand, for
link-targeted attacks the most effective appear to be the highest link degree
targeted attacks according to the recalculated (PTN of Berlin, Dallas, D\"usseldorf,
Istanbul, London, Moscow, Paris) or initial (Rome, Sa\~opaolo)
lists of links. Only for the PTN of Hamburg, Hong Kong, Taipei, and Sydney
the highest betweenness centrality
scenario appears to be the most effective, however even in this case the difference between
different scenarios is not much pronounced.
This similarity in behaviour for 'initial' and 'recalculated'
scenarios seems to be an intrinsic feature of the link-targeted attacks. Moreover, as
we noticed before, sometimes the 'initial' approach occurs to be more effective.
It is interesting to mention
that for three PTN (Hamburg, Istanbul, Sydney) which are not very
resilient against any kind of attacks (however not for PTN of
Dallas, which is least), most efficient is the scenario of removing
links with initial highest values of the betweenness centrality
${\cal C}_{B}^{i,(l)}$. It is worthwhile to note here, that the order of the PTN according
to their vulnerability under link-targeted attacks is similar to
that for the node-targeted scenarios, there are just few light
shifts.

\begin{figure}[th]
\centerline{\includegraphics[width=5.5cm]{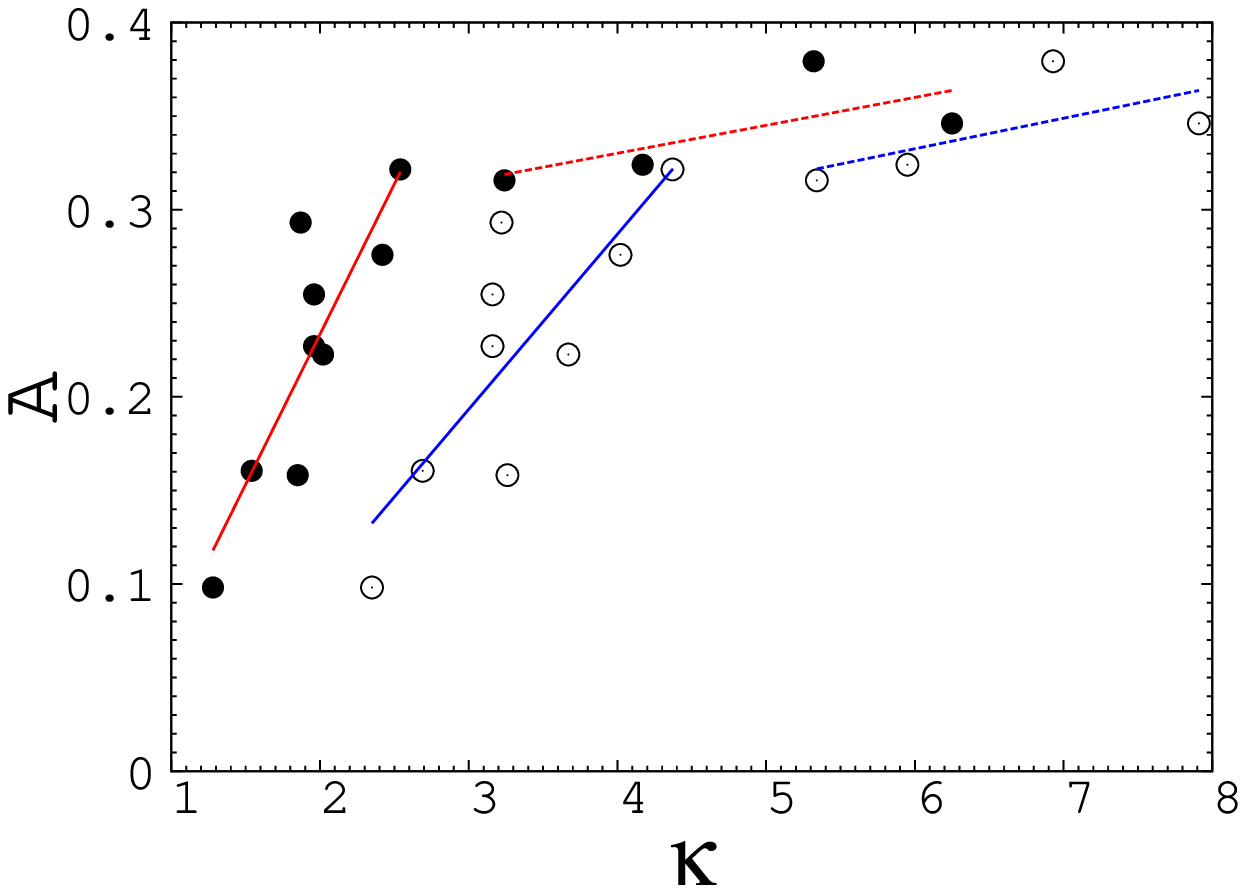} \hspace{3em}
\includegraphics[width=5.5cm]{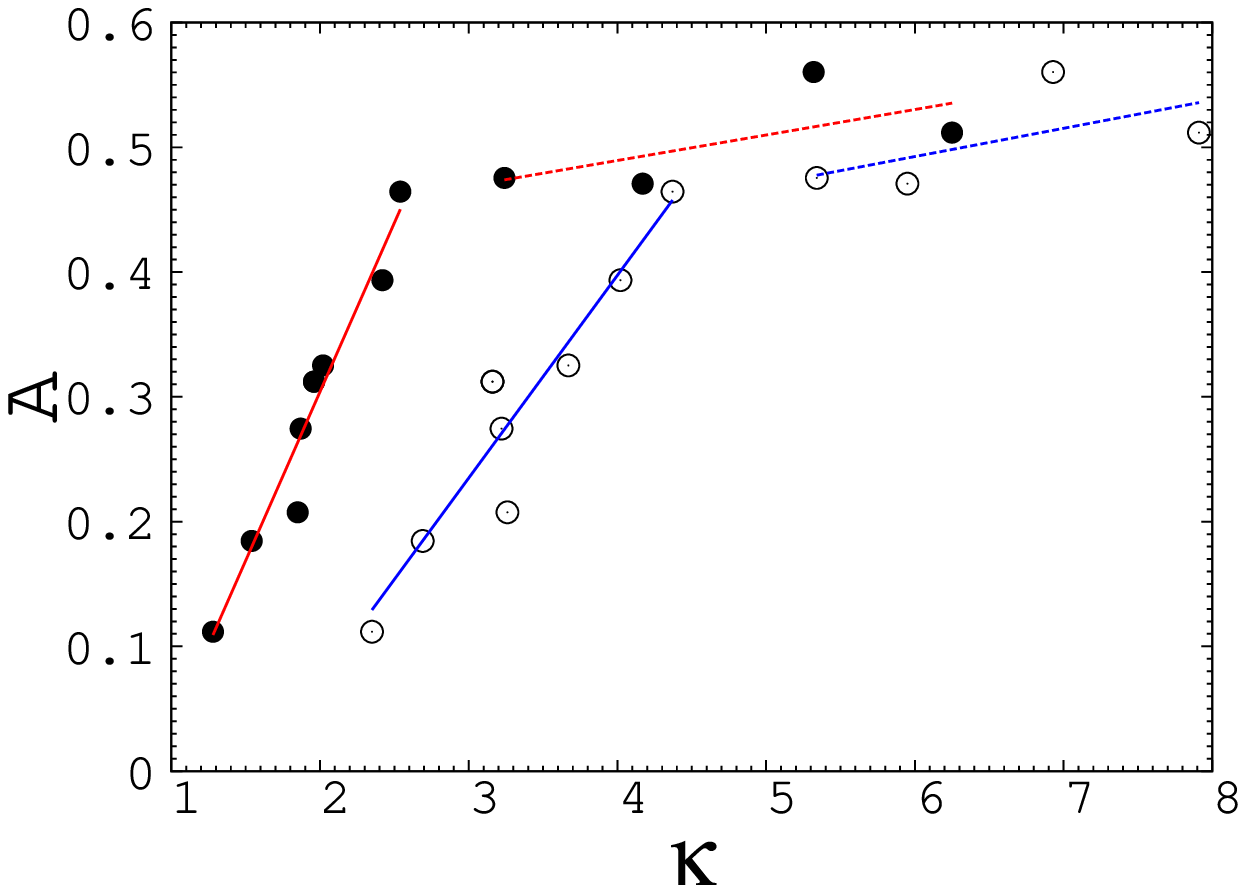}} \centerline{{\bf a}
\hspace{21em} {\bf b}}
\centerline{\includegraphics[width=5.5cm]{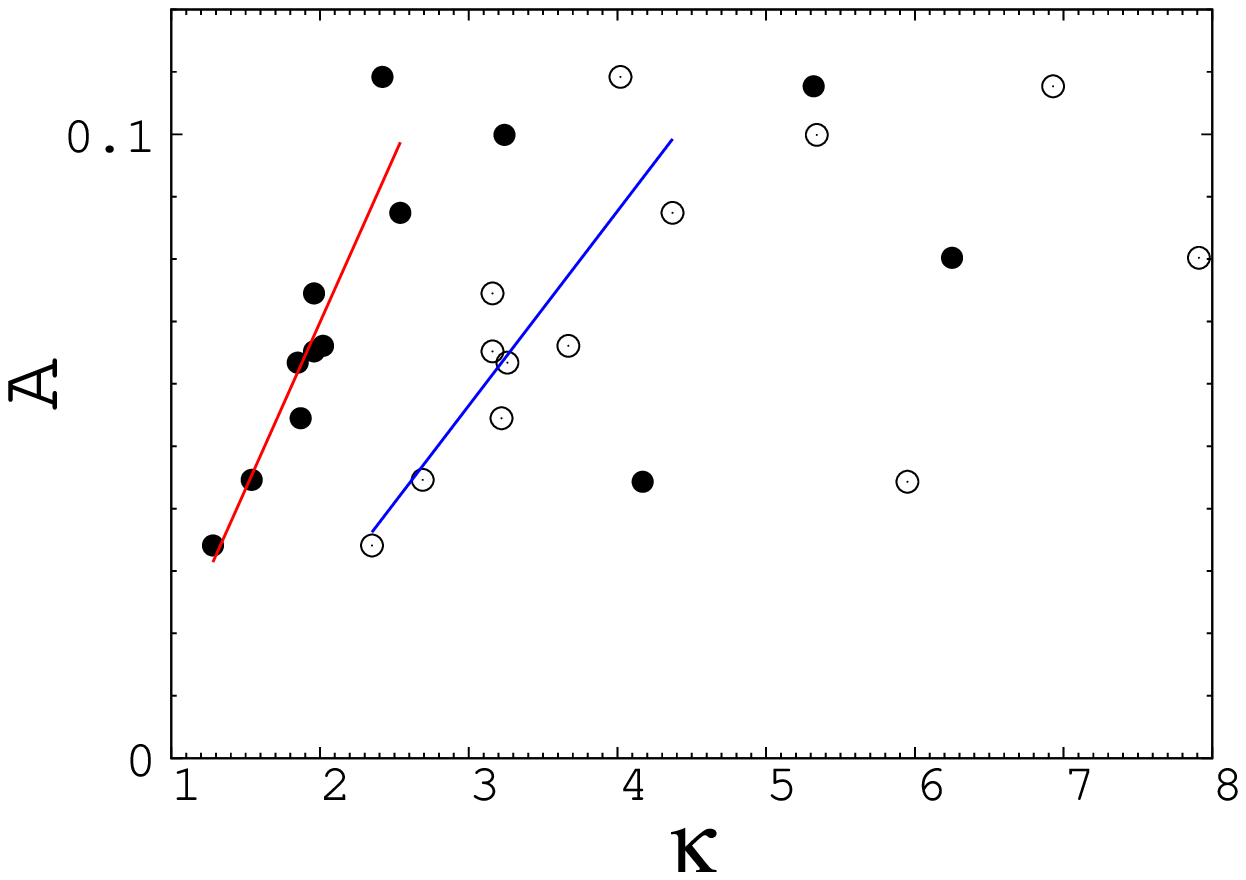} \hspace{3em}
\includegraphics[width=5.5cm]{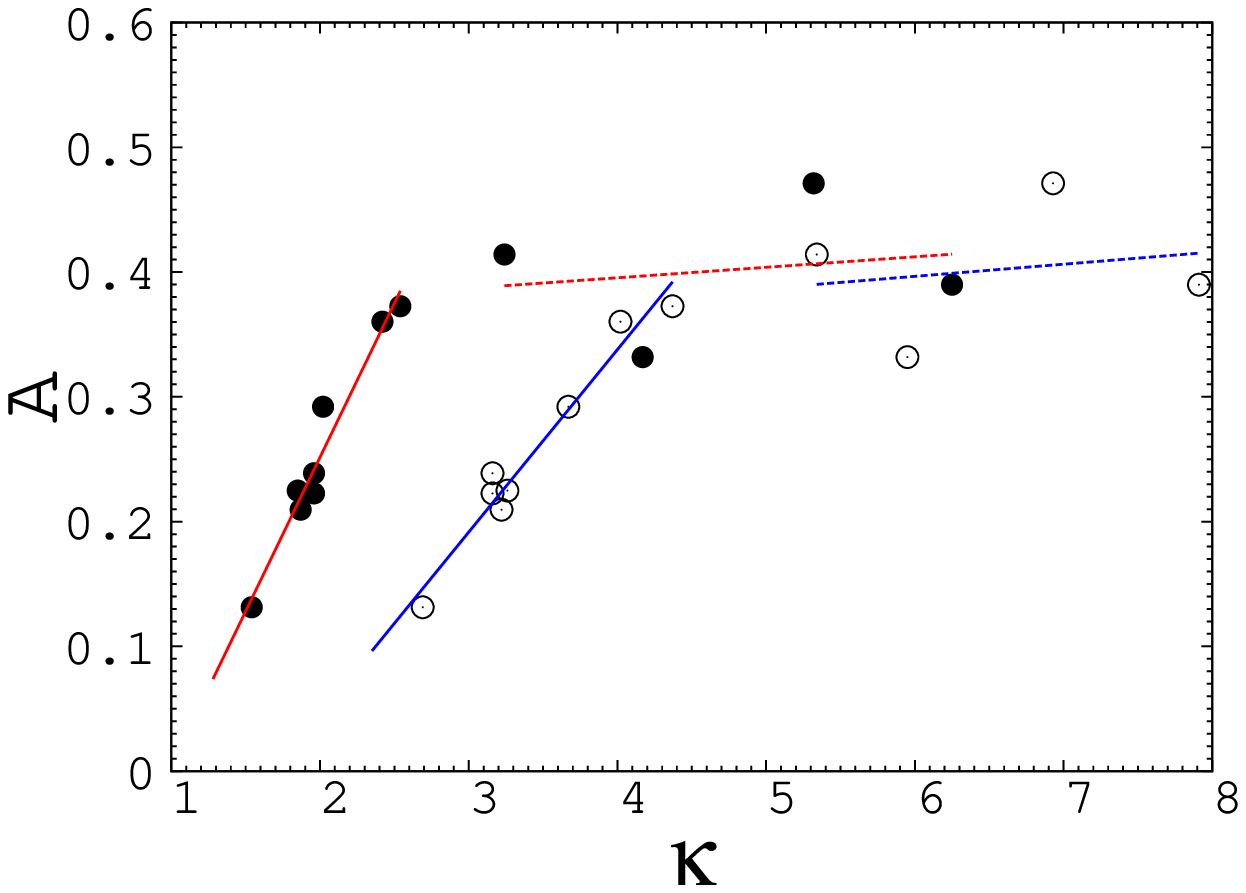}} \centerline{{\bf c}
\hspace{21em} {\bf d}}
\centerline{\includegraphics[width=5.5cm]{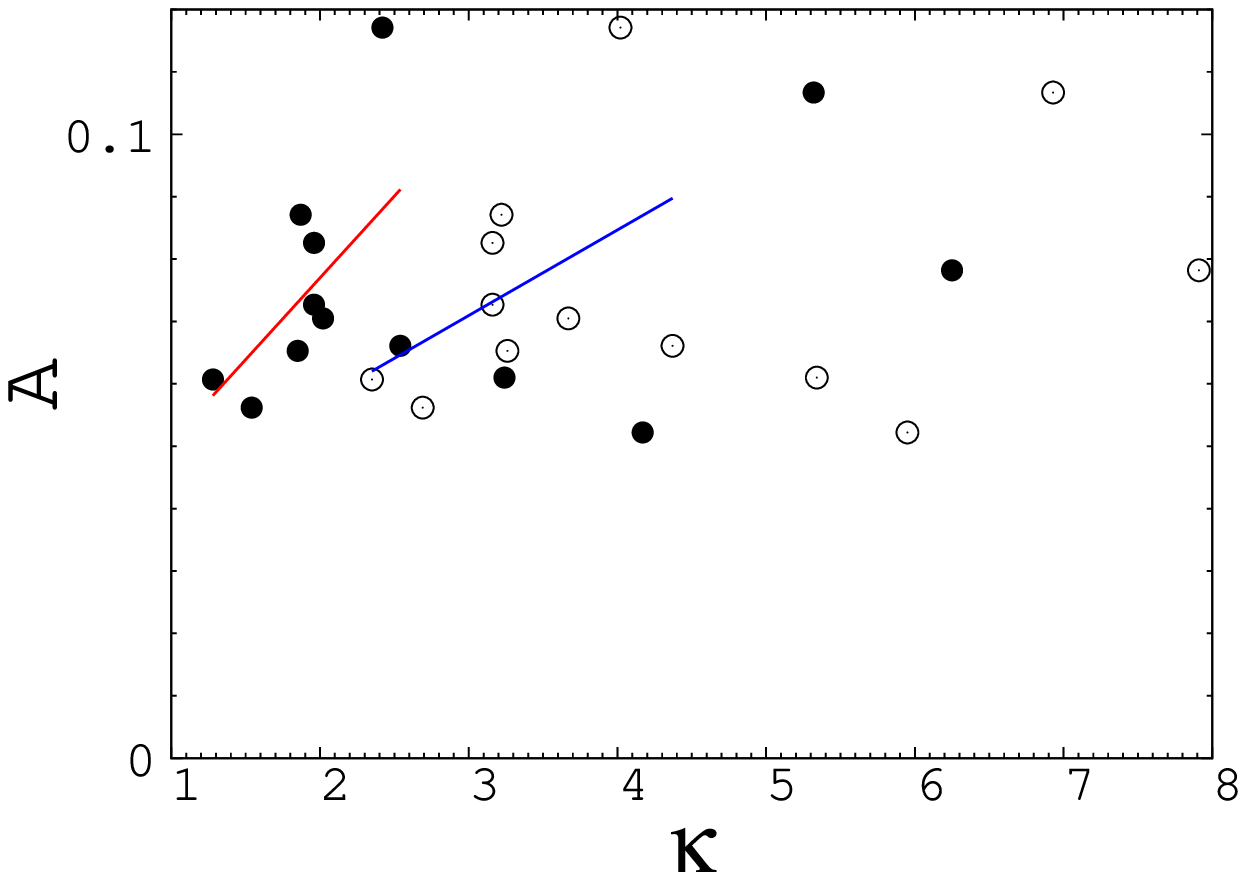} \hspace{3em}
\includegraphics[width=5.5cm]{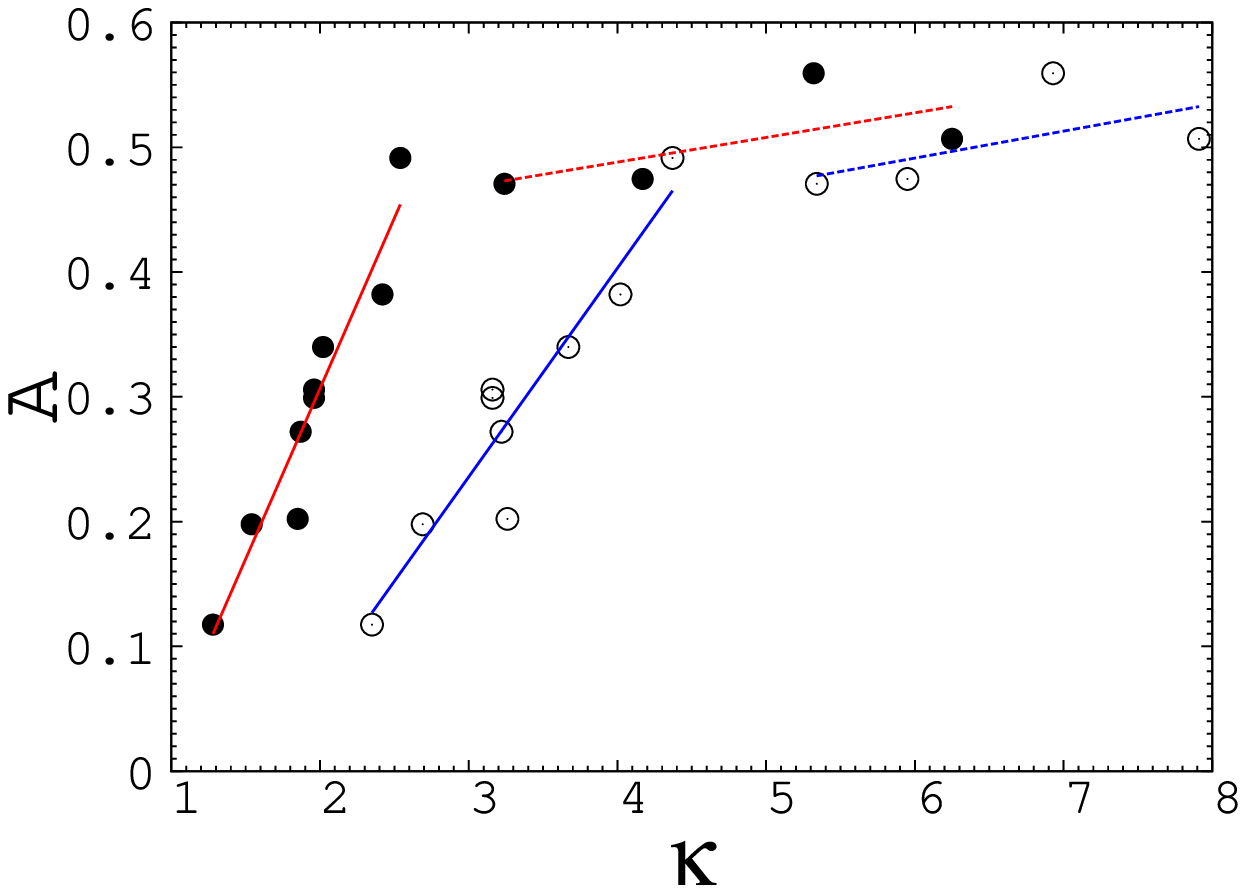}} \centerline{{\bf e}
\hspace{21em} {\bf f}}
 \vspace*{8pt}
\caption{Attacks on nodes (left column) and on links (right column).
Correlation between $A$ and $\kappa$ for the random ({\bf a, b}),
recalculated node degree ({\bf c, d}) and recalculated betweenness
({\bf e, f}) scenarios. Results for $\kappa^{(z)}$ are shown by
filled circles, results for $\kappa^{(k)}$ are shown by open
circles. Solid lines show linear fits of the corresponding data
points. \label{fig5}}
\end{figure}

To further shed light on correlation between the network
characteristics {\em prior to the attack} and their stability {\em
during the attack} we check correlation of $A$, Eq. (\ref{5a}), for
all PTN out of our database at different attack scenarios (table
\ref{tab2}) with the  value of the Molloy-Reed parameters
$\kappa^{(k)}$, Eq. (\ref{2}), and $\kappa^{(z)}$, Eq. (\ref{4}) of
the unperturbed networks, as given in Table \ref{tab1}. The results
are displayed in Figs. \ref{fig5}. There we show the value of $A$
correlated with the Molloy-Reed parameters $\kappa^{(z)}$ (filled
circles), and $\kappa^{(k)}$ (open circles) of the same network for
the node- and link-targeted attacks (left and right columns,
correspondingly). One notices two different regimes in the behavior
of the relation between $A$ and $\kappa$ for random and recalculated
highest degree scenarios both for node- and link-targeted attacks.
First, $A$ rapidly increases with an increase of $\kappa$, then, in
the second regime, when $\kappa$ exceeds certain 'marginal' value,
there is no pronounced correlation between $A$ and $\kappa$ any
more, however still a weak increase of $A$ with $\kappa$ is
observed.  These two regimes are observed both in $A(\kappa^{(z)})$
and $A(\kappa^{(k)})$ functions, however the behavior is more
pronounced in $A(\kappa^{(z)})$ plots (filled circles). We show the
linear fits for both regimes by solid lines in the figures. The
region of $\kappa$ where the first regime is observed is $1 \lesssim
\kappa^{(z)} \lesssim 2$ ($2 \lesssim \kappa^{(k)} \lesssim 4$).
Thus, if two PTNs  have initial values of corresponding Molloy-Reed
parameters in this region, it is very probable, that the PTN with
higher value of $\kappa$ will be essentially more stable than the
PTN with lower value of $\kappa$. However, the PTNs with the Molloy
Reed parameters  $\kappa^{(z)}
> 2$ ($\kappa^{(k)} > 4$) although in general being more stable than
those with lower $\kappa$ do not differ substantially in their
stability. A similar bahaviour is observed for the link-targeted
highest betweenness centrality attacks (Fig. \ref{fig5} {\bf f}) but
it is less pronounced, even less pronounced it is for the
node-targeted highest betweenness centrality attacks (Fig.
\ref{fig5} {\bf e}), where almost no correlation between $A$ and
$\kappa$ is observed. To understand the origin of the particular
sensitivity of PTN stability for small values of $\kappa$, let us
recall the results for uncorrelated networks (see formulas
(\ref{3}), (\ref{5}) and references in the text): a GCC in an
infinite network can exist only if $\kappa$ exceeds the marginal
value of $\kappa^{(z)} =1$ ($\kappa^{(k)} =2$). In the vicinity of
this marginal value the network is especially sensitive to even
slight changes. Obviously, the finiteness of the PTN and the
correlation effects present there lead to a variation for the
criteria (\ref{3}), (\ref{5}), however a general sensitivity of
network stability to the changes in $\kappa$ for small $\kappa$
remains.

\begin{figure}[th]
\centerline{\includegraphics[width=5.5cm]{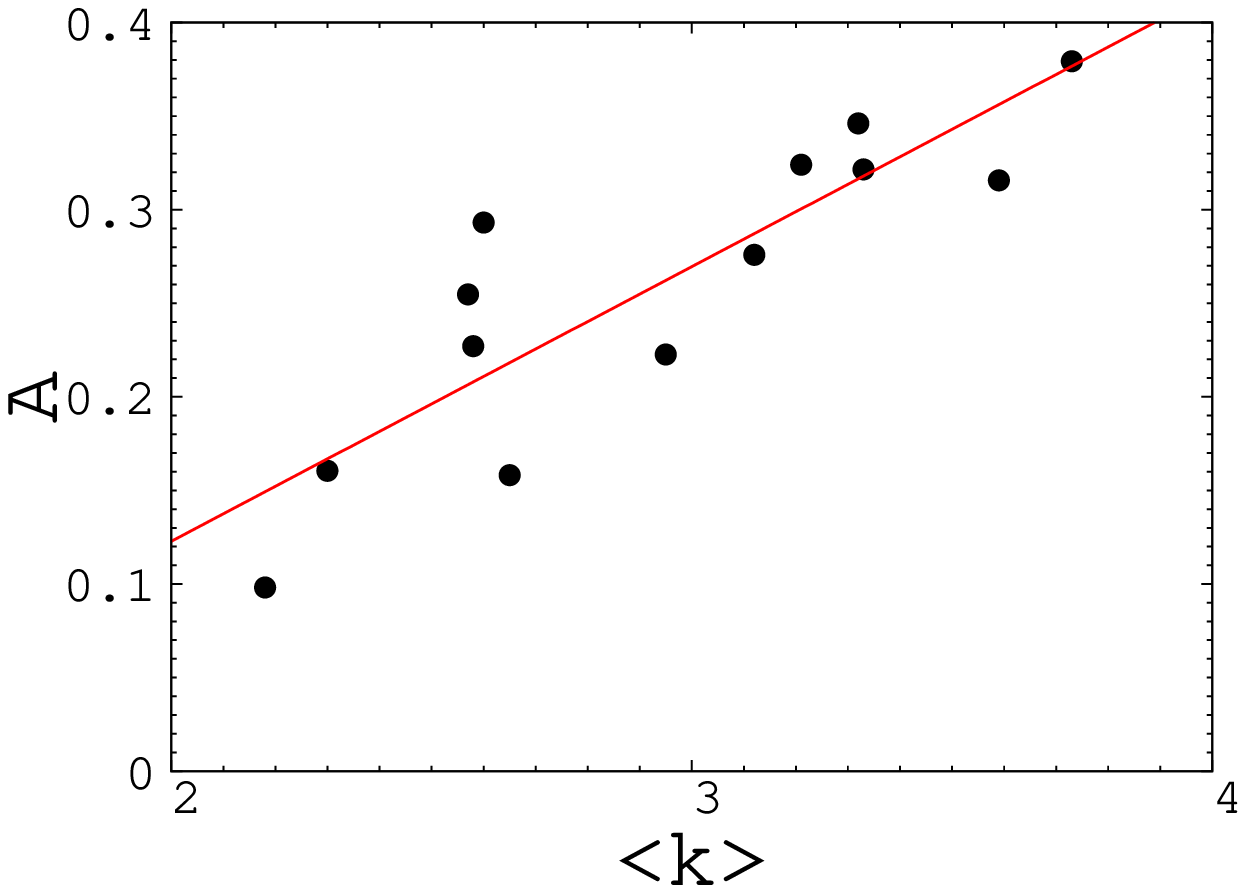} \hspace{3em}
\includegraphics[width=5.5cm]{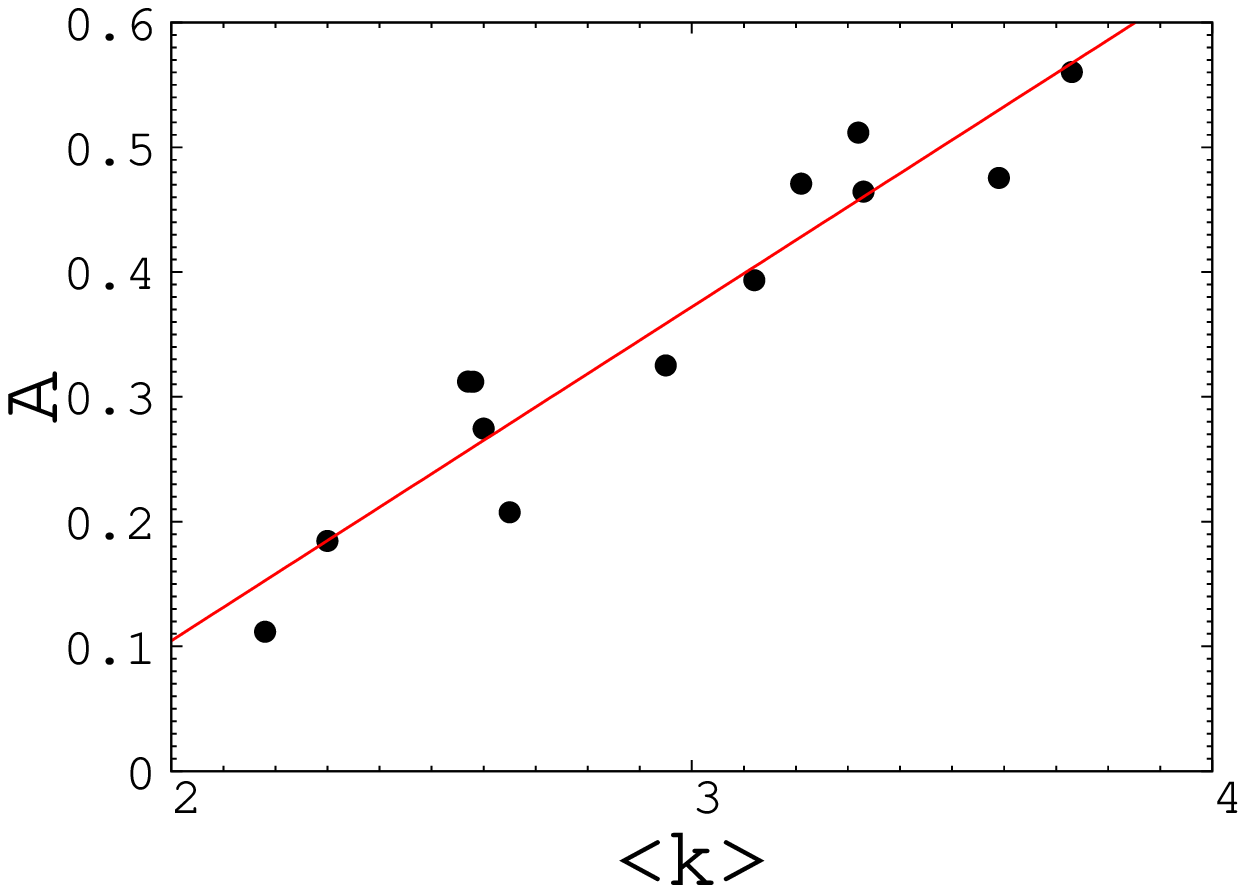}} \centerline{{\bf a}
\hspace{21em} {\bf b}}
\centerline{\includegraphics[width=5.5cm]{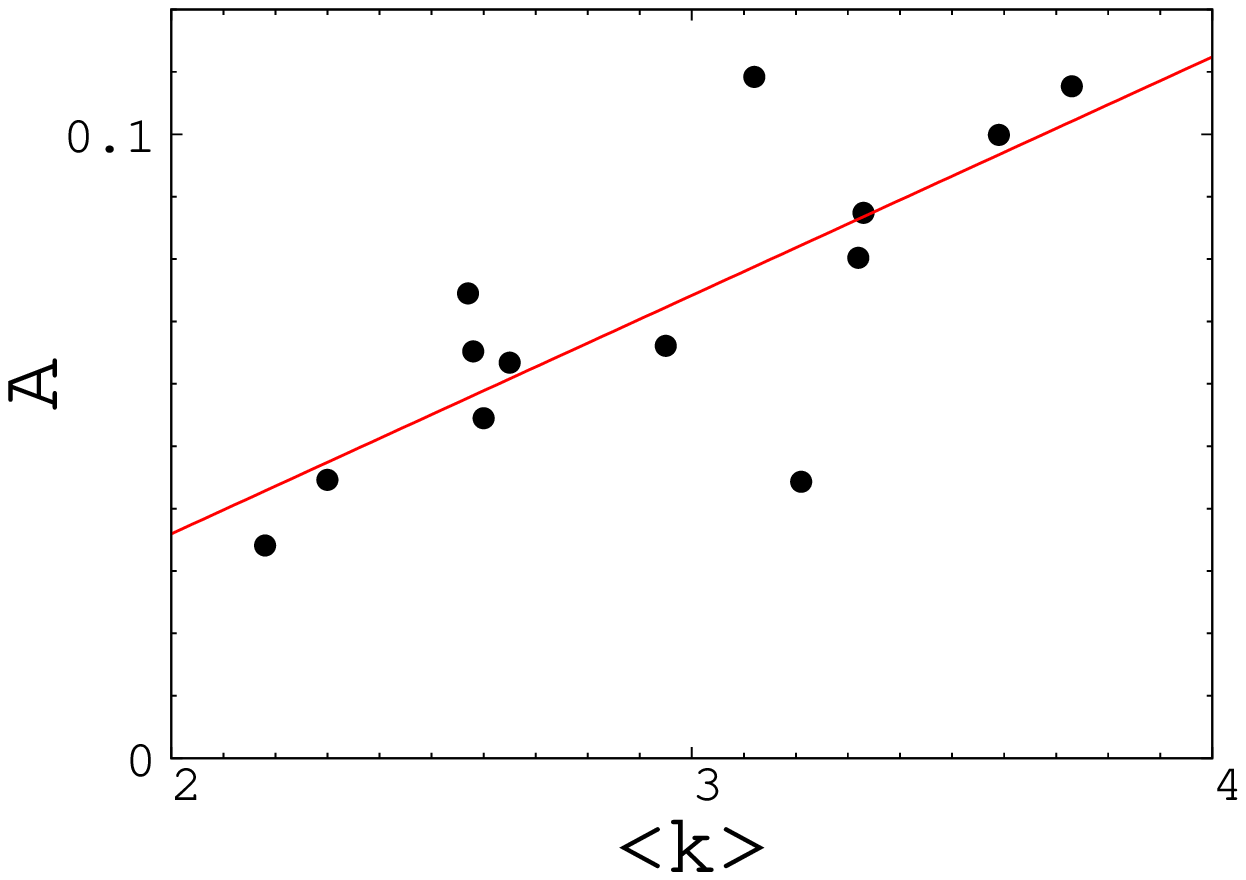}
\hspace{3em} \includegraphics[width=5.5cm]{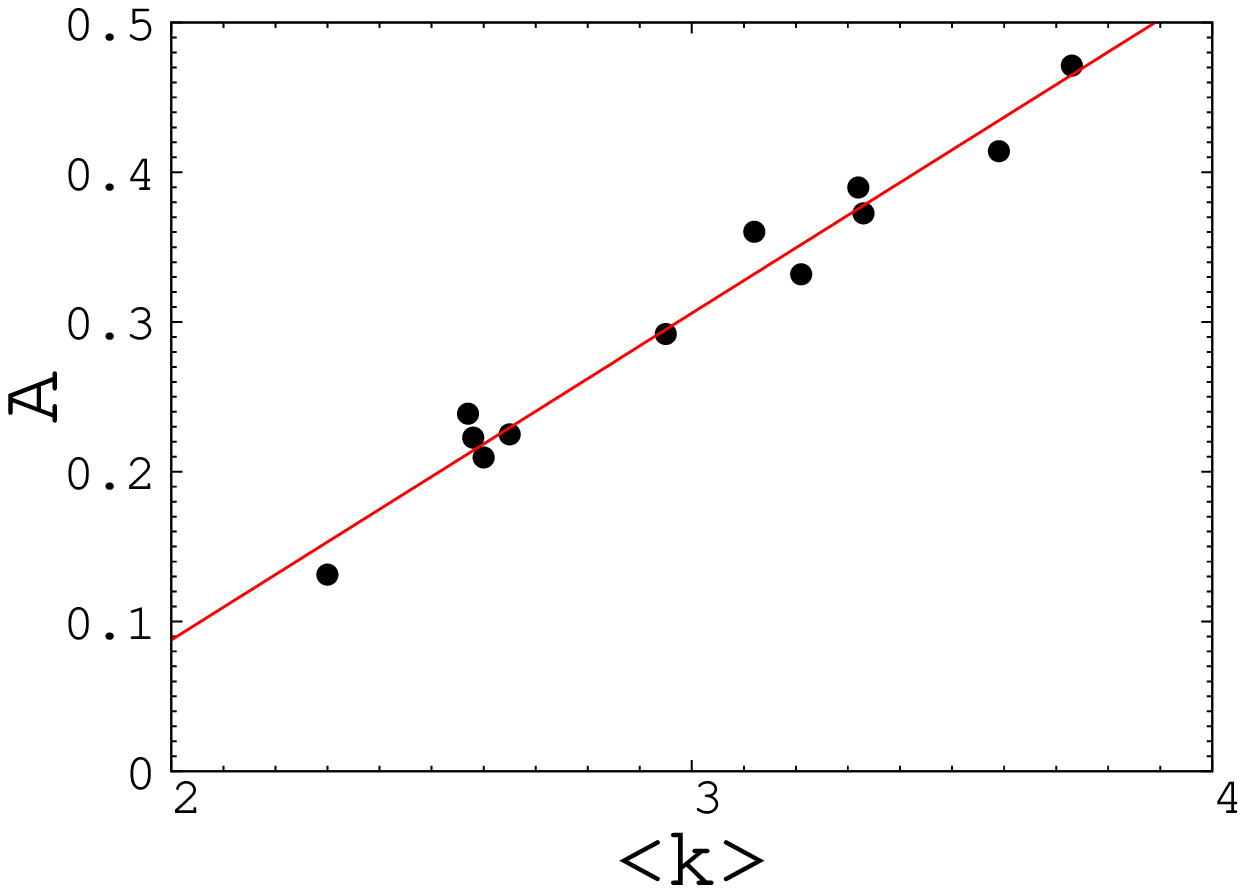}}
\centerline{{\bf c} \hspace{21em} {\bf d}}
\centerline{\includegraphics[width=5.5cm]{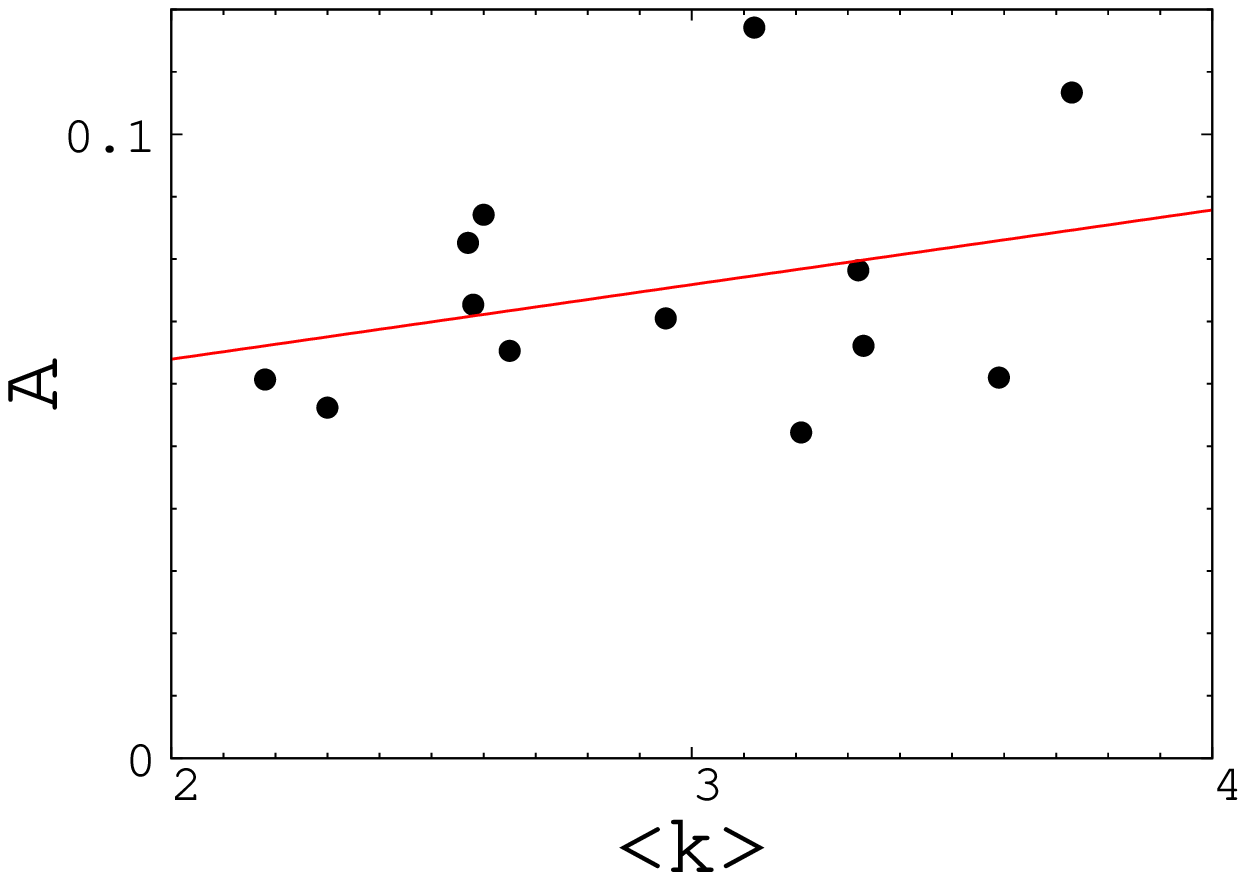} \hspace{3em}
\includegraphics[width=5.5cm]{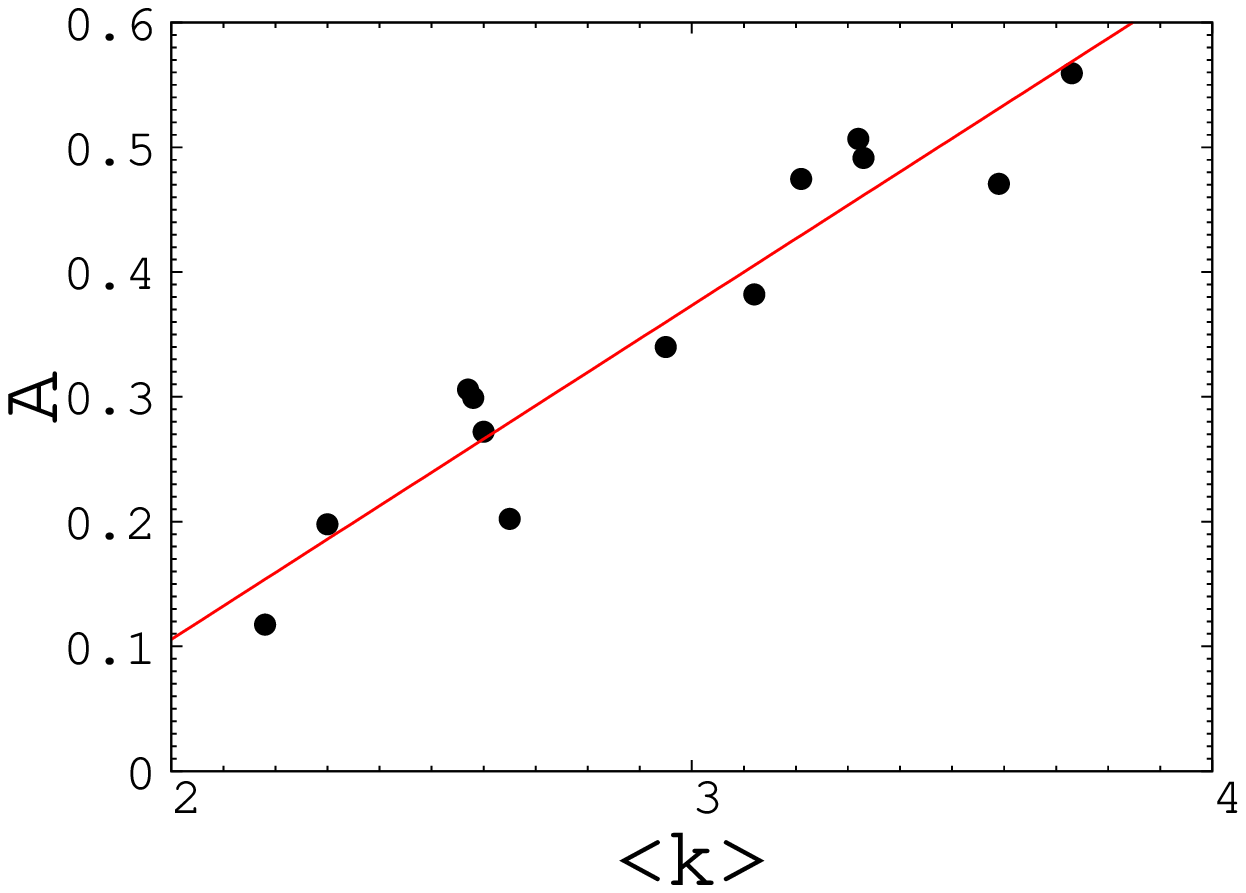}} \centerline{{\bf e}
\hspace{21em} {\bf f}}
 \vspace*{8pt}
\caption{Results of node targeted attacks (left column) and link
targeted attacks (right column). Correlation between $A$ and
$\langle k \rangle$ for  ({\bf a, b}) the random,  ({\bf c, d}) the
recalculated degree and  ({\bf e, f}) the recalculated betweenness
scenarios. \label{fig6}}
\end{figure}

Another interesting observation is illustrated by Figs. \ref{fig6}.
There, we show the correlation of $A$ with the mean node degree
$\langle k \rangle$ for the random ({\bf a, b}), recalculated degree
({\bf c, d}) and recalculated betweenness ({\bf e, f}) scenarios. A
generic feature of the $A(\langle k \rangle)$ plots is the linear
increase of $A$ with increasing of $\langle k \rangle$ which is
observed for all values of $\langle k \rangle$ and for all three
scenarios.  A similar increase is observed both for the node- and
link-targeted attacks, c.f. Figs. \ref{fig6} ({\bf a, c}) and Figs.
\ref{fig6} ({\bf b, d}), however the linear approximation holds for
the node-targeted attacks with less accuracy and is almost useless
for the highest betweenness centrality plots, Fig. \ref{fig6} ({\bf
e}). The corresponding fits are shown by solid lines in the figures.
The plots of Figs. \ref{fig6} demonstrate correlation of the network
stability with the initial 'density' of network constituents, nodes
or links, without relation to the correlations in the PTN structure.
This is different to the plots of Fig. \ref{fig5}, where the
correlations where considered by analyzing the second moment of the
node degree distribution $\langle k^2 \rangle$, that enters the
Molloy-Reed parameter. Therefore, Fig. \ref{fig6} shows the
correlation of the network stability measure $A$ with the mean node
degree,
 $\langle k\rangle$. There, for both cases, within the expected scatter of data
one observes clear evidence of an increase of $A$ with $\langle k
\rangle$, i.e. networks with smaller mean node degree $\langle k
\rangle$ break down at smaller values of $c$ and are thus more
vulnerable to the attacks. Again, this observation holds for the
link-targeted attacks as well for the node-targeted attack of random
and recalculated highest degree scenarios.

\begin{figure}[th]
\centerline{\includegraphics[width=5.5cm]{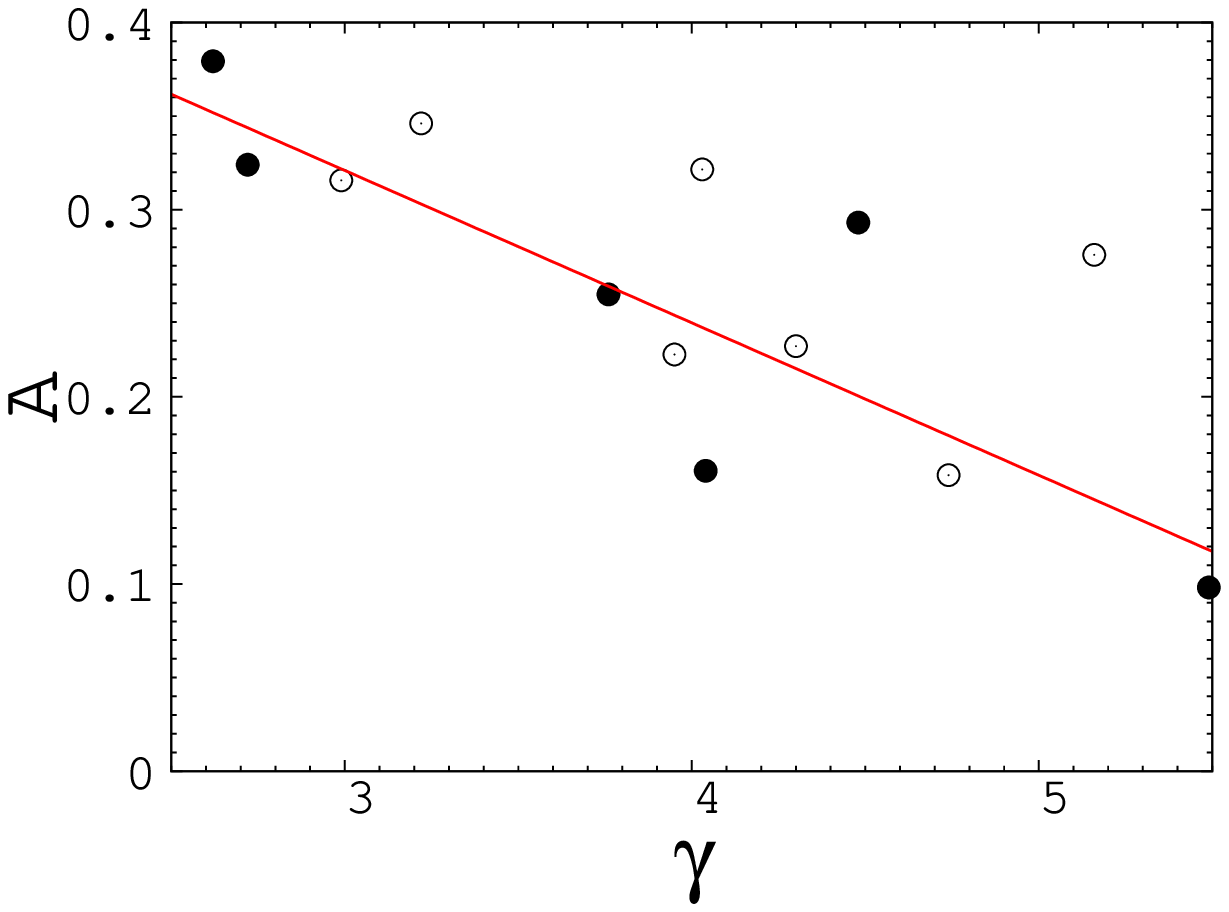} \hspace{3em}
\includegraphics[width=5.5cm]{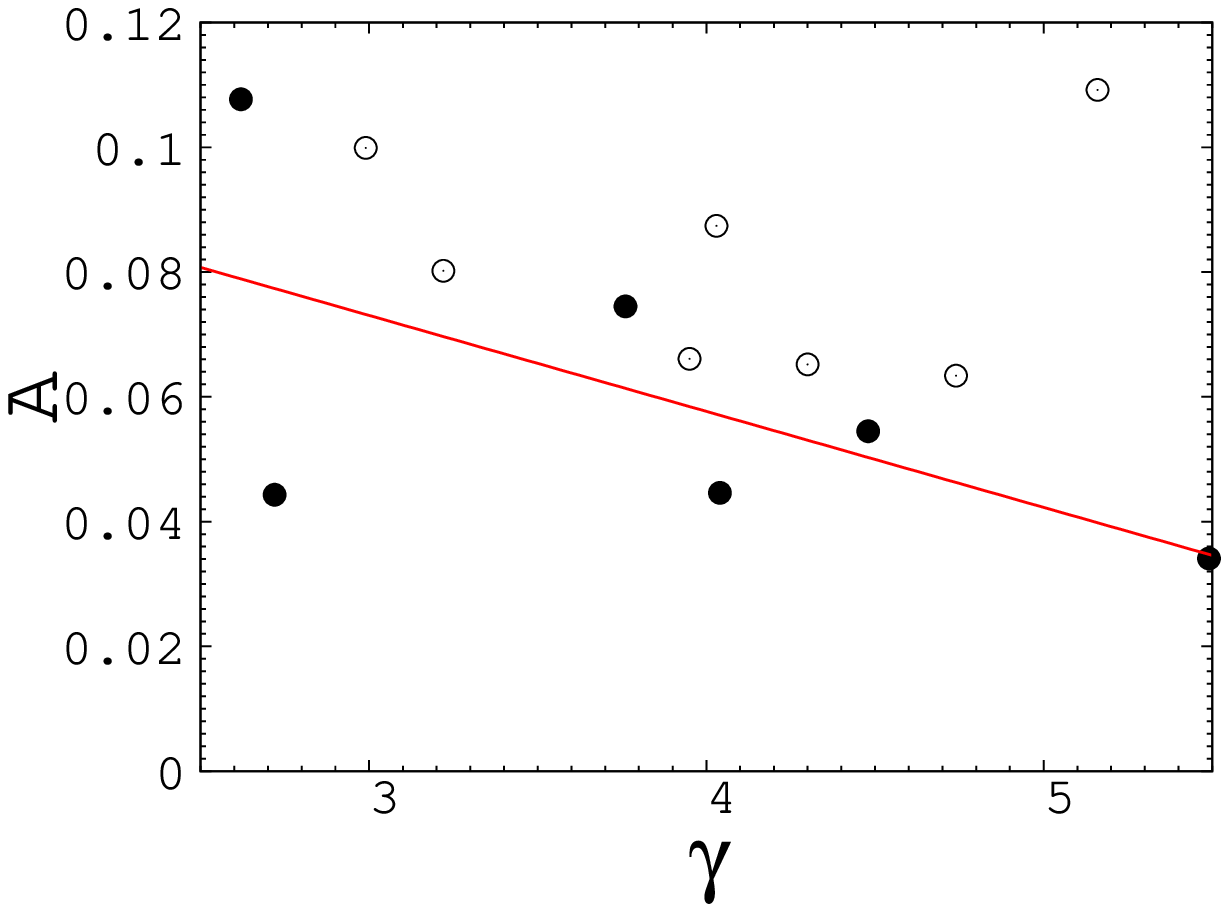}} \centerline{{\bf
a} \hspace{21em} {\bf b}}
 \vspace*{8pt}
\caption{Results of node targeted attacks. Correlation of $A$ with
respect to $\gamma$ for  ({\bf a}) the random and  ({\bf b}) the
recalculated node degree scenarios. Filled circles correspond to the
PTNs with more pronounced power-law decay of the node-degree
distribution, open circles correspond to the PTNs where the
power-law decay is less pronounced (see the section \ref{II}).
\label{fig7}}
\end{figure}

For the node-targeted attacks on scale-free networks it is useful
also to check the correlation between the node degree distribution
exponent $\gamma$, Eq. (\ref{6}) and the stability measure $A$.
Analytic results for infinite scale-free networks as well as
empirical observations for numerous real-world scale-free networks
have confirmed a particular stability of scale-free networks: there
is no percolation threshold for exponents $\gamma\leq 3$
\cite{Cohen00,Callaway00}. As we have observed in the previous
studies \cite{Ferber09a} some of the PTNs under consideration are
scale-free: their node-degree distributions have been fitted to a
power-law decay (\ref{6}) with the exponents shown in Table
\ref{tab1}. Others are characterized rather by an exponential decay,
but up to a certain accuracy they can also be approximated by a
power-law behavior (then, the corresponding exponent is shown in
Table \ref{tab1} in brackets). In Fig. \ref{fig7}{\bf a} and
\ref{fig7}{\bf b} we show the correlation between the fitted
node-degree distribution exponent $\gamma$ and $A$ for the random
and recalculated node degree scenarios.  One observes a notable
tendency to find PTNs with smaller values of $\gamma$ to be more
resilient as indicated by larger values of $A$. This tendency holds
even if we include the PTNs which are better described by the
exponential decay of the node-degree distributions.

\section{Conclusions and outlook} \label{VI}
In this paper we have presented an empirical analysis of the
reaction of PTNs of different cities of the world upon random failure or
directed attack scenarios.
There may be numerous reasons for individual failure, ranging from a
random accident to a targeted destruction. However, in
accumulation these may lead to an emergent behavior as a result of which
the PTN ceases to function. On the one hand our analysis is motivated by
practical interest in the stability of individual PTNs thereby
comparing the operating features of different PTNs. On the
other hand we were seeking to identify criteria, which  allow to judge {\em a
priori} on the attack stability of real world correlated networks
of finite size checking how do these criteria correspond to the
analytic results available for the infinite uncorrelated networks.

To perform the present analysis we have used previously accumulated
\cite{Ferber09a} data on PTNs of several major cities of the world
(see table \ref{tab1}) and simulated attacks of different scenarios
targeted on the PTN nodes and links. To quantify the  PTN stability
to attacks of different scenarios we use a recently introduced
\cite{schneider11a,schneider11b} numerical measure of network
robustness. In our case, this measure is defined as the area below
curve described by the normalized size $S(c)$ of the largest connected
component as function of the share $c$ of removed nodes.
In this respect, the measure captures the overall resilient behavior
over the complete attack sequence. Table
\ref{tab2} allows to compare the robustness of a given PTN to attacks of
different scenarios as well as to compare the relative robustness of
different PTNs.

The comparison of PTN characteristics measured {\em prior} to the attack
with the PTN robustness monitoring its behaviour {\em during} the
attack allowed us to propose criteria that allow an a priori
estimate of PTN robustness and stability with respect to an attack.
This stability is indicated by a high value of the Molloy-Reed
parameters $\kappa^{(k)}$, Eq. (\ref{2}), and $\kappa^{(z)}$, Eq.
(\ref{4})  as well as by the high value of the mean node
degree $\langle k \rangle$ of the unperturbed networks. Moreover, if
the PTN node degree distribution manifests a power-law decay, we
have observed a notable tendency to find PTNs with smaller values of
$\gamma$ to be more stable.

\section*{Acknowledgments}
BB, CvF, and YuH acknowledge partial support by the FP7 EU IRSES
project N269139 'Dynamics and Cooperative Phenomena in Complex
Physical and Biological Media'. This work was in part performed in
frames of the COST Action MP0801 'Physics of Competition and
Conflicts'.

\end{document}